\documentclass[a4paper,fleqn,usenatbib]{mnras}
\usepackage{newtxtext,newtxmath}
\usepackage[dvipsnames]{xcolor}
\usepackage{ae,aecompl}
\usepackage{graphicx}	
\usepackage{amsmath}	
\usepackage{mathtools}
\usepackage{array}
\usepackage{booktabs}

\usepackage{hyperref}
\hypersetup{draft=false}

\title[DEVILS: The sSFR-M$_{\star}$ plane part I]{Deep Extragalactic VIsible Legacy Survey (DEVILS):  The sSFR-M$_{\star}$ plane part I: The recent SFH of galaxies and movement through the plane}
\author[L. J. M. Davies]{L. J. M. Davies$^{1}$\thanks{E-mail:
 luke.j.davies@uwa.edu.au},  J. E. Thorne$^{1}$, S. Bellstedt$^{1}$, R. H. W. Cook$^{1}$, M. Bravo$^{2}$,  \newauthor A. S. G. Robotham$^{1,3}$,  C. del P. Lagos$^{1,3,4}$, S. Phillipps$^{5}$,  M. Siudek$^{6,7}$, B. W. Holwerda$^{8}$, \newauthor M. N. Bremer$^{5}$,  J. D'Silva$^{1}$,  and S. P. Driver$^{1}$ \\
 $^{1}$ ICRAR, The University of Western Australia, 35 Stirling Highway, Crawley, WA 6009, Australia \\
$^{2}$ Department of Physics \& Astronomy, McMaster University, 1280 Main Street W, Hamilton, ON, L8S 4M1, Canada \\
$^{3}$ ARC Centre of Excellence for All Sky Astrophysics in 3 Dimensions (ASTRO 3D) \\
$^{4}$ Cosmic Dawn Center (DAWN), Denmark \\
$^{5}$ Astrophysics Group, School of Physics, University of Bristol, Bristol BS8 1TL, UK \\
$^{6}$ Institute of Space Sciences (ICE, CSIC), Campus UAB, Carrer de Can Magrans, s/n, 08193 Barcelona, Spain \\
$^{7}$Institut de Física d’Altes Energies (IFAE), The Barcelona Institute of Science and Technology, 08193 Bellaterra (Barcelona), Spain\\
$^{8}$ Physics \& Astronomy Department, University of Louisville, Louisville, KY 40292, USA \\
}

\date{Accepted XXX. Received YYY; in original form ZZZ}

\pubyear{2025}

\begin{document}
\label{firstpage}
\pagerange{\pageref{firstpage}--\pageref{lastpage}}
\maketitle

\begin{abstract}
In a recent paper we parameterised the evolution of the star-formation rate dispersion ($\sigma_{SFR}$) across the specific star-formation rate - stellar mass plane (sSFR-M$_{\star}$) using the Deep Extragalactic VIsible Legacy Survey (DEVILS) - suggesting that the point at which the minimum in the dispersion occurs (M$^{*}_{\sigma-min}$) defines a boundary between different physical mechanisms affecting galaxy evolution. Here we expand upon that work to determine the movement of galaxies through the sSFR-M$_{\star}$ plane using their recent star-formation histories (SFHs) and explore how this leads to the observed $\sigma_{SFR}$-M$_{\star}$ relation. We find that galaxies in sub-regions of the sSFR-M$_{\star}$ plane show distinctly different SFHs, leading to a complex evolution of the sSFR-M$_{\star}$ plane and star-forming sequence (SFS). However, we find  that selecting galaxies based on stellar mass and position relative to SFS alone (as is traditionally the case), may not identify sources with common recent SFHs, and therefore propose a new selection methodology. We then use the recent SFH of galaxies to measure the evolution of the SFS, showing that it has varying contributions from galaxies with different SFHs that lead to the observed changes in slope, normalisation and turnover stellar mass. Finally, we determine the overall evolution of the sSFR-M$_{\star}$ plane from $z\sim1$ to today. In the second paper in this series we will discuss physical properties of galaxies with common recent SFHs and how these lead to the observed $\sigma_{SFR}$-M$_{\star}$ relation and evolution of the sSFR-M$_{\star}$ plane.

\end{abstract}


\section{Introduction}

The star-forming galaxy `main' sequence (SFS), and more broadly the specific star-formation rate - stellar mass (sSFR-M$_{\star}$) plane, is now regarded as a fundamental parameter space in terms of our understanding of the evolution of galaxies \citep[$e.g.$][]{Daddi07, Elbaz07,Noeske07,Salim07,Whitaker12, Speagle14, Johnston15, Davies16b} \textcolor{black}{and potentially more fundamentally as a derived state from a pressure-regulated feedback-modulated model of star-formation \citep[$e.g.$][]{Lin19, Ellison20, Ellison24} }. The position of a galaxy within this plane \textcolor{black}{can} encode a huge amount of information regarding its past evolution (\textcolor{black}{such as,} the stellar mass formed in situ and accumulated through mergers to date) and future evolution (star formation and to some extent the remaining gas reservoir assuming a scaling relation). \textcolor{black}{However, care must be taken when directly relating a galaxy's position within the plane to physical properties, as multiple correlated properties can ultimately lead to an individual galaxy's location \citep[$e.g.$][]{Abramson16}}. Despite this, the parameter space and the position of galaxies within it is fertile ground for exploring the various astrophysical properties that shape galaxy evolution processes, such as Active Galactic Nuclei (AGN) feedback \citep{Kauffmann04,Fabian12}, stellar feedback \citep{Dekel86, DallaVecchia08, Scannapieco08}, mergers \citep[$e.g.$][]{Bundy04, Baugh06, Kartaltepe07, Bundy09,Jogee09,deRavel09,Lotz11,Robotham14}, morphological/structure evolution \citep{Conselice14, Eales15}, gas fuelling \citep{Kauffmann06,Sancisi08, Mitchell16}, secular quenching \citep{Schawinski14, Barro13}, environmental quenching \citep[$e.g.$][]{Giovanelli85, Peng10, Cortese11, Darvish16, Davies19b}, etc. This has led to the SFR-M$_{\star}$ plane and its populations being used as a key diagnostic tool in many studies. These works predominately use the position of galaxies within the SFR-M$_{\star}$ plane with respect to the SFS to identify galaxies that are either currently, or at some point in the past, undergoing evolutionary processes that cause them to deviate from the SFS locus. By this logic the SFS is thought to contain `typical' star-forming galaxies that exist in a self-regulated state \citep[$i.e.$][]{Bouche10, Daddi10, Genzel10, Lagos11, Lilly13, Dave13, Mitchell16, Wang21}, where the inflow rate of gas for future star-formation is balanced by the rate at which new stars are formed and the outflow of gas from feedback events ($i.e.$ Supernovae, SNe, and AGN). Populations that deviate from this self-regulated state are thought to have undergone astrophysical processes that cause them to leave the locus of the SFS. The goal of this area of galaxy evolution science is to constrain the key processes that lead to these deviations, typically by identifying commonalities in populations selected with respect to the SFS. 

However, this picture is a complicated one. The processes that cause galaxies to leave the locus of the SFS likely vary as a function of stellar mass, morphology, structure, environment, available gas supply and the presence of an AGN. In addition, the SFS itself is not a static entity and evolves with time \citep[e.g.][]{Lee15, Thorne21}; likely in a manner that is not self-similar across stellar masses and environments. As such, even defining a galaxy's position relative to the SFS, or what that means in terms of its current star formation, is fraught with difficulty; let alone picking apart the dominant astrophysical mechanisms that caused an individual galaxy to leave this relation. Consequently, this has been a very active field of research over the last decade or more.

To first order, the current position of a galaxy within this plane is largely governed by its star-formation history \citep[SFH,][]{Madau98, Kauffmann03, Carnall19, Bellstedt20}. This not only encodes the stellar mass formed to date and the current star-formation rate, but importantly can also be used to map a galaxy's trajectory through the SFR-M$_{\star}$ plane, $i.e.$ we can essentially determine its stellar mass and star-formation rate at any given epoch. \textcolor{black}{For example, the works of \cite{Sanchez18, Iyer18, Ciesla17, Ciesla21} and others have used this approach to great effect in both tracking galaxies through the plane and aiming to determine the root causes of galaxy's trajectories.}  However, care must be taken as such approach ignores the impact of mergers ($i.e.$ it assume all stellar material at the observation epoch has been in the same system throughout its lifetime). 

In addition, any method for determining a galaxy's SFH is likely constrained by i) the stellar population, metallicity evolution,  dust and AGN models used, ii) the fitting procedure and iii) the time resolution that can be obtained from the input observational data - where we note that errors on an individual galaxy's SFH typically get larger the further we extrapolate into that galaxy's past (to first order, it is easier to differentiate between blue young stars and red old stars, than red old stars and red very old stars). Despite this, using SFHs to track a galaxy's path through the SFR-M$_{\star}$ plane can yield key results regarding the formation and evolutionary trends affecting the fundamental characteristics of the observed SFR-M$_{\star}$ plane, such as the SFS, star-burst population \citep[$e.g.$][]{Broussard19, Emami19}, passive population, `green valley' galaxies \citep[$e.g.$][]{Bremer18,Phillipps19}, and turnover population at high stellar masses \citep[point there the SFS become non-linear, $e.g.$ see][]{Whitaker14, Schreiber15, Thorne21}.          

\textcolor{black}{A different but complementary approach to using the full SFH to track the trajectory of galaxies through the SFR-M$_{\star}$ plane, specifically in terms of star-bursting systems, uses the ratio of luminosity/SFRs in different star-formation tracers, which probe different timescales. For example, H$\alpha$ emission from nebular gas can be traced back to only the hottest and youngest stars, and varies on short timescales directly with the current very young stellar population, while emission in $e.g.$ the ultra-violet (UV) or mid-infrared (MIR) can be traced to a broader age range of stars, and can lag behind any recent short-duration changes to star-formation \citep[e.g. see][for discussion of star-formation rate timescales from different observational tracers]{Davies15b, Davies16b}. This can be used to explore recent short-timescale changes to star-formation activity (bursting/quenching) by comparing $e.g.$  H$\alpha$ to UV luminosities \citep[][]{Broussard19, Emami19}.} 

\textcolor{black}{These approaches have also been tested on high resolution hydrodynamical simulations, and have been shown to accurately reproduce the recent SFHs of galaxies \citep[$e.g$][]{Flores21}, and have been used to account for recently discovered populations of highly star-forming systems in the early Universe \citep[$e.g$][]{Sun23, Gelli23, Asada24}. These approaches have also been used to show that a galaxy's location relative to the SFS may not be indicative of it recent SFH \citep[$e.g.$][]{Asada24}.  However, robust H$\alpha$ measurements can be limited for large samples of galaxies at appreciable redshifts due to the observational limitations in obtaining the required high signal-to-noise spectroscopic measurements. This can limit the applicability of these approaches to large sample of galaxies. For example, within the sample presented in this work using the Deep Extragalactic VIsible Legacy Survey \citep[DEVILS, ][]{Davies18, Davies21} we do not obtain high enough signal-to-noise spectroscopic observations to explore the H$\alpha$/UV ratio as a tracer of recent SFH. Even in the more local Galaxy And Mass Assembly \citep[GAMA,][]{Driver11,Hopkins13,Liske15,Driver16, Baldry18} sample our H$\alpha$ measurements are likely not robust enough to perform such an analysis, and the same will be true for the Wide-Area VISTA Extragalactic Survey \citep[WAVES,][]{Driver19}. However, for these samples we will have robust SED-derived SFHs, which we focus on in this work. }    

\vspace{2mm}

A somewhat crude, but quantifiable, metric of the deviation of galaxies away from the locus of the SFS and the variability of SFHs at a given stellar mass is the star-formation rate dispersion - stellar mass relation ($\sigma_{\mathrm{SFR}}$-M$_{\star}$). This relation is a measure of the spread of current SFRs at a given stellar mass about the SFS locus, where larger spreads suggest stronger astrophysical processes that drive galaxies away from the self-regulated star-formation model \textcolor{black}{\citep{Kurczynski16, Boogaard18, Davies19b, Davies22, Cole23}}. In \cite{Davies19b} we explored the local $\sigma_{\mathrm{SFR}}$-M$_{\star}$ in GAMA. We found that the relation follows a `U-shaped' distribution, with large dispersion at both high and low stellar masses and a low dispersion region at around log$_{10}$(M/M$^{*})$=9-10 \citep[consistent with other observational work, $e.g.$][]{Willett15}. We also found that this relation is ubiquitous irrespective of selection method and SFR indicator. Following this, in \cite{Davies22}, hereafter D22, we expanded upon the GAMA analysis to explore the evolution of the $\sigma_{\mathrm{SFR}}$-M$_{\star}$ relation using DEVILS. Here we found that the `U-shaped' $\sigma_{\mathrm{SFR}}$-M$_{\star}$ relation persists to intermediate redshifts, and that the minimum SFR dispersion point, M$^{*}_{\sigma-min}$, moves to higher stellar mass with redshift \textcolor{black}{- noting that this holds true for both directly measured SFR dispersions and the intrinsic dispersion, accounting for observational errors/uncertainties}.  We also found that the M$^{*}_{\sigma-min}$ point occurs at the same specific star-formation rate (sSFR) at all epochs, and argue that the M$^{*}_{\sigma-min}$, and the sSFR at which it occurs, is a key fundamental property in understanding the evolution of the SFR-M$_{\star}$ plane.    

A potential physical interpretation of these results is that the large dispersion at the low stellar mass end is caused by stochastic star-formation processes leading to both star-burst and quenching events, and driving galaxies away from the locus of the SFS. At high stellar masses AGN feedback leads to quenching events, pushing galaxies below the SFS and \textcolor{black}{potentially} increasing the dispersion. Between these two regimes galaxies are massive enough to not be strongly affected by stochastic star-formation, and also not massive enough to harbour powerful AGN, leading to a low dispersion about the SFS. This model is given credence by the work of \cite{Katsianis19}, and others, who find a `U-shaped' $\sigma_{\mathrm{SFR}}$-M$_{\star}$ relation in the EAGLE hydrodynamical simulations. They then re-run the simulations with stochastic star-formation turned off (removing the scatter at low stellar masses) and AGN feedback turned off (removing the scatter at high stellar masses). 

This however, is just one proposed set of physical mechanisms that can explain the $\sigma_{\mathrm{SFR}}$-M$_{\star}$ relation, and many other factors could be at play, such as the impact of long-duration SFH variation, environmental quenching mechanisms, and/or morphological and structural evolution.  Our previous observational works have largely focused on quantifying the shape and evolution of the $\sigma_{\mathrm{SFR}}$-M$_{\star}$ relation, and showing that the observed observational trends are largely consistent with the physical interpretation presented in \cite{Katsianis19}. They do not in themselves provide strong observational evidence of the true underlying astrophysics that results in the shape of the $\sigma_{\mathrm{SFR}}$-M$_{\star}$ relation, or allow us to differentiate between different physical interpretations. Here, and in the second paper in this series, we expand upon these works with the aim of constraining the astrophysical processes that lead to the distribution of galaxies in the sSFR-M$_{\star}$ plane, and the $\sigma_{\mathrm{SFR}}$-M$_{\star}$ relation and its evolution. We explore the recent SFH of galaxies in different regions of the sSFR-M$_{\star}$ plane, to show how this important parameter space is evolving with time. 

\vspace{2mm}

In the first paper in this series we will explore the variation of SFHs across the sSFR-M$_{\star}$ plane and determine how this leads to the overall evolution of the SFS. In this paper we largely focus on the sample selection, observational trends and how they shape the observed evolution of the sSFR-M$_{\star}$ plane.  In the second paper we explore the potential astrophysical drivers of the variation of SFHs across the plane, which ultimately lead to these observed distributions. Paper I is structured as follows: In Section \ref{sec:select}, we select galaxies with potential common recent SFHs, based on their position within the sSFR-M$_{\star}$ plane with respect to stellar mass and the SFS (as is traditionally the case), and show that there is strong variation in SFHs both in terms of sSFR and stellar mass (Section \ref{sec:var}). However, we also show that this traditional method for selecting galaxies with common SFHs may be flawed, as galaxies selected based on their position relative to the SFS can show a broad range of SFHs, particularly at high stellar masses. We then define a metric for parametrising the recent SFH of a galaxy (Section \ref{sec:deltSFR}) and use this to identify new selection regions which isolate common SFHs (Section \ref{sec:newreg}). Finally, we explore the evolution of the SFS and SFR-M$_{\star}$ in terms of populations with common SFHs, to highlight how the varying contributions of these population as a function of time leads to the observed evolution of the shape and normalisation of the SFS (Section \ref{sec:planeEvol}). In paper II we will expand upon this to explore the physical properties of galaxies with common SFHs, identified here.

\section{Data and Sample Selection}

\subsection{The Deep Extragalactic VIsible Legacy Survey}

DEVILS is a spectroscopic survey undertaken at the Anglo-Australian Telescope (AAT), which aimed to build a high completeness ($>$85\%) sample of $\sim$50,000 galaxies to Y$<$21\,mag in three well-studied deep extragalactic fields: D10 (COSMOS), D02 (ECDFS) and D03 (XMM-LSS). The survey will provide the first high completeness sample at $0.3<z<1.0$, allowing for the robust parametrisation of group and pair environments in the distant Universe. The science goals of the project are varied, from the environmental impact on galaxy evolution at intermediate redshift, to the evolution of the halo mass function over the last $\sim$7\,billion years. For full details of the survey science goals, survey design, target selection, photometry and spectroscopic observations see \cite{Davies18, Davies21}.

The DEVILS regions were chosen to cover areas with extensive exisiting and oncoming imaging to facilitate a broad range of science. In this work we only use the DEVILS D10 field which covers the Cosmic Evolution Survey region \citep[COSMOS,][]{Scoville07}, extending over 1.5deg$^{2}$ of the UltraVISTA \citep{McCracken12} field and centred at R.A.=150.04, Dec=2.22. This field is prioritised for early science as it is the most spectroscopically complete, has the most extensive multi-wavelength coverage of the DEVILS fields, and has already been processed to derive robust galaxy properties through spectral energy distribution (SED) fitting.

\begin{figure}
\begin{center}
\includegraphics[scale=0.13]{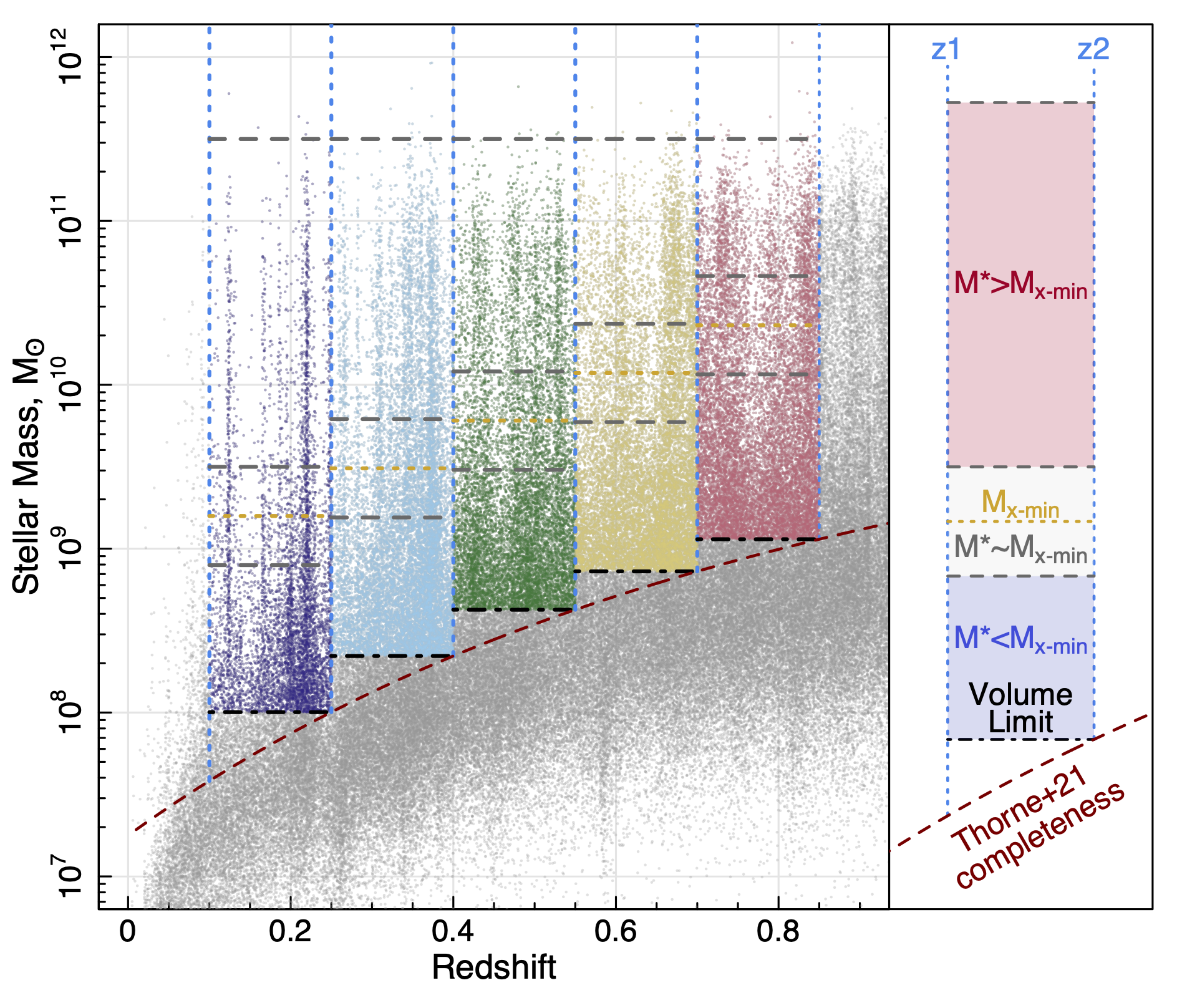}

\caption{The sample selection of galaxies used in this work, taken from the redshift - stellar mass distribution, M$_{\star}(z)$. The left panel displays our various samples at each redshift, while the right panel describes all of the various lines and regions on the left panel at a single epoch. The grey points display the full DEVILS D10 sample taken from \citet{Thorne21}. First, following \citet{Davies22} we split our sample into five $\Delta z=0.15$ redshift bins at $0.1<z<0.85$, divided by the dashed vertical blue lines. Next we display the colour-completeness limit as a function of redshift taken from \citet{Thorne21} as the dashed red line. To define a volume-limited sample at each epoch, we take the \citet{Thorne21}  colour completeness point at the upper redshift end of each redshift bin (black dot-dashed horizontal lines) and only include galaxies above this limit at each epoch. This results in the coloured points, and forms our samples for further analysis. On this figure we then also show the minimum SFR dispersion point M$^{*}_{\sigma-min}$ at each epoch, taken from \citet{Davies22}, as the dashed horizontal gold lines. Later in this work we will select galaxies in three stellar mass ranges with respect to M$^{*}_{\sigma-min}$, which are bounded by the grey dashed horizontal lines.  As noted, all of these lines/regions are summarised in the right panel.            }
\label{fig:Mz}
\end{center}
\end{figure}

In this work we use the outputs of the SED fitting process outlined in \cite{Thorne21} and \cite{Thorne22}. Briefly, \cite{Thorne21} fits galaxies in the D10 region using the \textsc{ProSpect} \citep{Robotham20} SED fitting code to estimate galaxy properties such as stellar mass, SFR, SFH and metallicity. In \cite{Thorne22}, this processes is updated to include an AGN model, which allows for the identification of sources hosting bright AGN and improvements to the other derived properties for AGN host galaxies. While the overwhelming majority of sources in the D10 sample do not change their properties, the sources identified as AGN do, in some case, have significant changes, particularly to their SFR and SFH (as UV and MIR-FIR is now attributed to the AGN and not star-formation). A detailed description of how galaxy properties in the D10 sample are affected by the inclusion of the AGN is included in \cite{Thorne22}, so we refer the reader to that work. \textcolor{black}{However, for completeness we briefly note that these works use photometric data described in \cite{Davies21} covering 22 bands from the FUV to FIR. As an example of the data quality, for the sample of galaxies used in this work at $0.4<z<0.55$ (see Figure \ref{fig:Mz}), the 5th percentile r-band signal-to-noise ratio is 4.5, with a median of signal-to-noise of 30.}

The methodology in both \textsc{ProSpect} analyses uses a parametric skew log-normal truncated SFH for all galaxies, which assumes that all SFHs are smoothly evolving (the limit of what \textcolor{black}{parametric SFHs fitted to} photometric data alone can provide). As such, in this work we will not be able to robustly explore very short duration burst-like SFH variation ($<$100\,Myr) \textcolor{black}{over the galaxy's lifetime.  However, this does not preclude this approach from identifying galaxies with recent busts of star-formation at $t\sim0$. Within the \textsc{ProSpect} analyses of \cite{Thorne22}, galaxies can be fit with a negative time peak in their log-normal SFH, such that their star-formation can rise rapidly to a peak at the observation epoch - which appears like a recent burst (these are observed in our sample discussed later in this work). The only caveat here, is that this process will not capture systems which have an old stellar population and a secondary recent burst of star-formation ($i.e.$ rejuvenated systems), where the log-normal SFH will fit the older population but fail to simultaneously fit the recent burst. However, in analysis of the \textsc{shark} semi-analytic model \citep{Lagos18, Lagos24}, these types of systems appear rare \citep[][]{Bravo22}. Despite this, in this work we do not claim that \textsc{ProSpect} is robust in exploring the SFH for individual sources but} instead we look for overall global changes to star-formation to the population. For example, while we may not capture the true SFH of a galaxy which is currently undergoing a \textcolor{black}{recent rejuvenated burst of star-formation, we may be able to identify that it is increasing in its overall star-formation activity and that this is linked to its position in the sSFR-M$_{\star}$ plane}. This is an important distinction, which will be readdressed later in this work. However, it is worth explicitly stating that while using \textsc{ProSpect}-derived SFHs means that we are unlikely to be able to robustly model the true SFH for all galaxies on short time-scales, we can identify average trends in the SFH of galaxies with common properties and in common regions of the sSFR-M$_{\star}$ plane. As such, in this work we limit ourselves to exploring commonalities in the SFHs of galaxies, not individual sources. In Appendix \ref{sec:shark} we explore differences between \textsc{ProSpect}-derived SFHs and true SFHs in the \textsc{shark} semi-analytic model \citep{Lagos18} and discuss the implications for our results.  We note that for this paper we use the DEVILS-internal D10-\textsc{ProSpect} catalogue \texttt{DR1\_v01}.

\vspace{2mm}

\textcolor{black}{Given the above caveats, it is also worth noting here that the \textsc{ProSpect} analyses from \cite{Thorne22} makes a (justified) \textit{choice} of using parametric SFHs in the fitting process (and other methodology choices). There is currently a wealth of ongoing literature regarding the the merits/failings of using parametric vs non-parametric SFHs \citep[$e.g.$][]{Tojeiro07, Dye08, Pacifici12, Pacifici16, Iyer19, Leja19} when fitting galaxy SEDs. Comparisons of these approaches are discussed extensively in many works and there is still ongoing debate as to the preferred approach \citep[$e.g.$][]{Carnall19,Lower20, Robotham20, Bellstedt20, Thorne21}. As such, we do not re-cover this extensively here.  However, one point we would like to highlight is that in the works of \cite{Carnall19} and \cite{Lower20}, they suggest that using parametric SFHs systematically biases measurements of stellar mass, SFR and recent SFH - which would impact our results derived here. However, these comparisons are with parametric SFHs using a delayed-Tau model \citep[akin to that used in the \textsc{magphys} SED fitting code,][]{daCunha08}. \textsc{ProSpect} does not use this functional form for its SFHs. In fact, in Figure 7 of \cite{Thorne21} it is shown that there is also systematic offset between \textsc{magphys}-derived and \textsc{ProSpect}-derived stellar masses and star-formation rates - similar to that found in \cite{Lower20}. \textcolor{black}{As such, the strongest systematic biases observed are likely a consequence of imposing a delayed-Tau model for SFHs, and may not be a generic issue with the use of all forms of parametric SFHs}. We also note, that not only the choice parametric vs non-parametric SFH, but also the choice of stellar population model, Initial Mass Function, dust prescription, inclusion/non-inclusion of an AGN model, metallically evolution, etc can also have a strong impact on SED-derived SFHs. For our DEVILS galaxies, the details of these choices and their motivation are described in \cite{Robotham20} and \cite{Thorne21}, and we simply use the output SFHs here. However, as above, we do aim to identify possible caveats in the interpretation of the results derived here, which may be influenced by the choice of SED fitting methodology.  }

\begin{figure*}
\begin{center}
\includegraphics[scale=0.6]{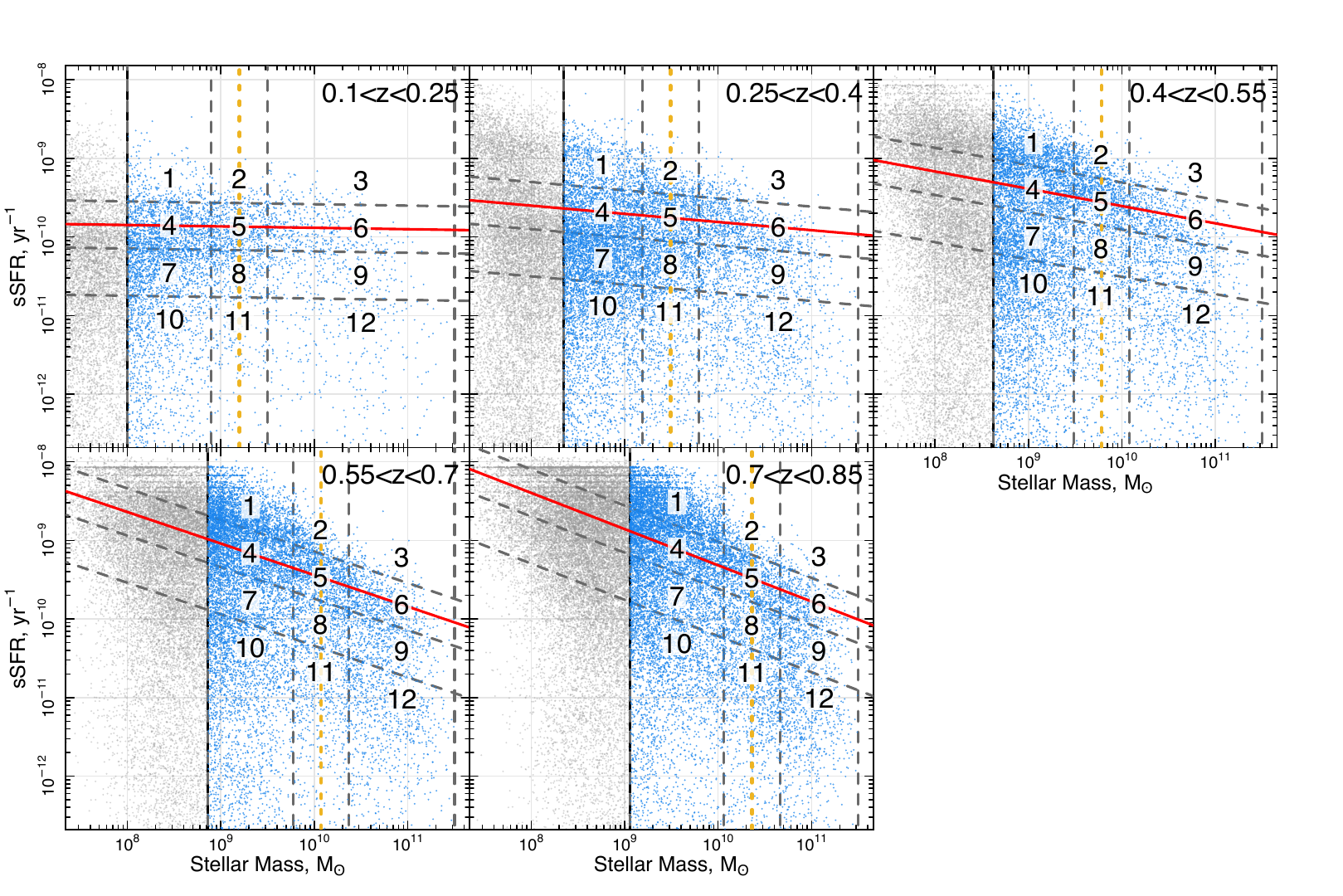}

\caption{The selection of different regions of the sSFR-M$_{\star}$ plane based on the the star-forming sequence (red line) and minimum SFR dispersion point, M$^{*}_{\sigma-min}(z)$ (vertical dashed gold line). At each epoch we define three stellar mass ranges based on M$^{*}_{\sigma-min}(z)$ at log$_{10}$(M$^{*}$)$<$log$_{10}$(M$^{*}_{\sigma-min}(z))$-0.3\,dex (regions 1, 4, 7, 10), log$_{10}$(M$^{*}_{\sigma-min}(z))$-0.3\,dex$<$log$_{10}$(M$^{*}$)$<$log$_{10}$(M$^{*}_{\sigma-min}(z))$+0.3\,dex (regions 2, 5, 8, 11), and  log$_{10}$(M$^{*}$)$>$log$_{10}$(M$^{*}_{\sigma-min}(z))$+0.3\,dex (regions 3, 6, 9, 12). We then also define four regions in sSFR in relation to the star-forming sequence (SFS, red line) at log$_{10}$(sSFR)$>$SFS+0.3\,dex (bursting, regions 1, 2, 3),  SFS-0.3\,dex$<$log$_{10}$(sSFR)$<$SFS+0.3\,dex (sequence, regions 4, 5, 6),    SFS-0.9\,dex$<$log$_{10}$(sSFR)$<$SFS-0.3\,dex (quenching, regions 7, 8, 9), and log$_{10}$(sSFR)$<$SFS-0.9\,dex (passive, regions 10, 11, 12). In combination these regions cover various parts of the sSFR-M$_{\star}$ plane where we may expect galaxies to be undergoing different physical process that are shaping their recent SFH. The stellar mass limit of our sample at each epoch is given by the dashed vertical black line, we do not consider systems below this point (grey points) as they are incomplete in colour and stellar mass.}
\label{fig:selection}
\end{center}
\end{figure*}

\subsection{Sample selection}
\label{sec:select}

Following the D22 analysis of the $\sigma_{\mathrm{SFR}}$-M$_{\star}$ relation, we also split our sample into five $\Delta z=0.15$ redshift bins between $0.1<z<0.85$.  These redshift bins are shown as the dashed vertical lines in Figure \ref{fig:Mz}.  Following this we only consider galaxies within a stellar mass, volume complete sample at each redshift bin. To determine this range, \cite{Thorne21} calculate the rest-frame $g-i$ colour completeness limit, $M_{\mathrm{lim}}$, as a function of stellar mass and look-back time as:

\begin{equation}
\mathrm{log}_{10}(M_{\mathrm{lim}}/M_{\odot})=\frac{1}{4}t_{\mathrm{lb}}+7.25
\end{equation}

\noindent where $t_{\mathrm{lb}}$ is look-back time in Gyrs. This line essentially represents the lower stellar mass at which the sample is complete to both red-passive and blue-starforming galaxies at a given epoch. Figure \ref{fig:Mz} displays this as the red dashed line. To form a colour-complete sample across each $\Delta z$ range, we calculate the \cite{Thorne21} completeness limit at the upper redshift end of each of our redshift samples, and only include galaxies above this stellar mass. These are then shown as the coloured points in Figure \ref{fig:Mz}. For reference, on Figure \ref{fig:Mz}, we also show the evolution of the minimum SFR dispersion point from D22,  M$^{*}_{\sigma-min}$($z$), as the horizontal gold line in each redshift bin. A range of M$^{*}_{\sigma-min}$($z$)$\pm$0.3\,dex (used later in this work) is shown as the grey horizontal lines that bound M$^{*}_{\sigma-min}$($z$). For the rest of this work we will show our analysis for each of the colour-complete $\Delta z$ ranges.

\begin{figure*}
\begin{center}
\includegraphics[scale=0.6]{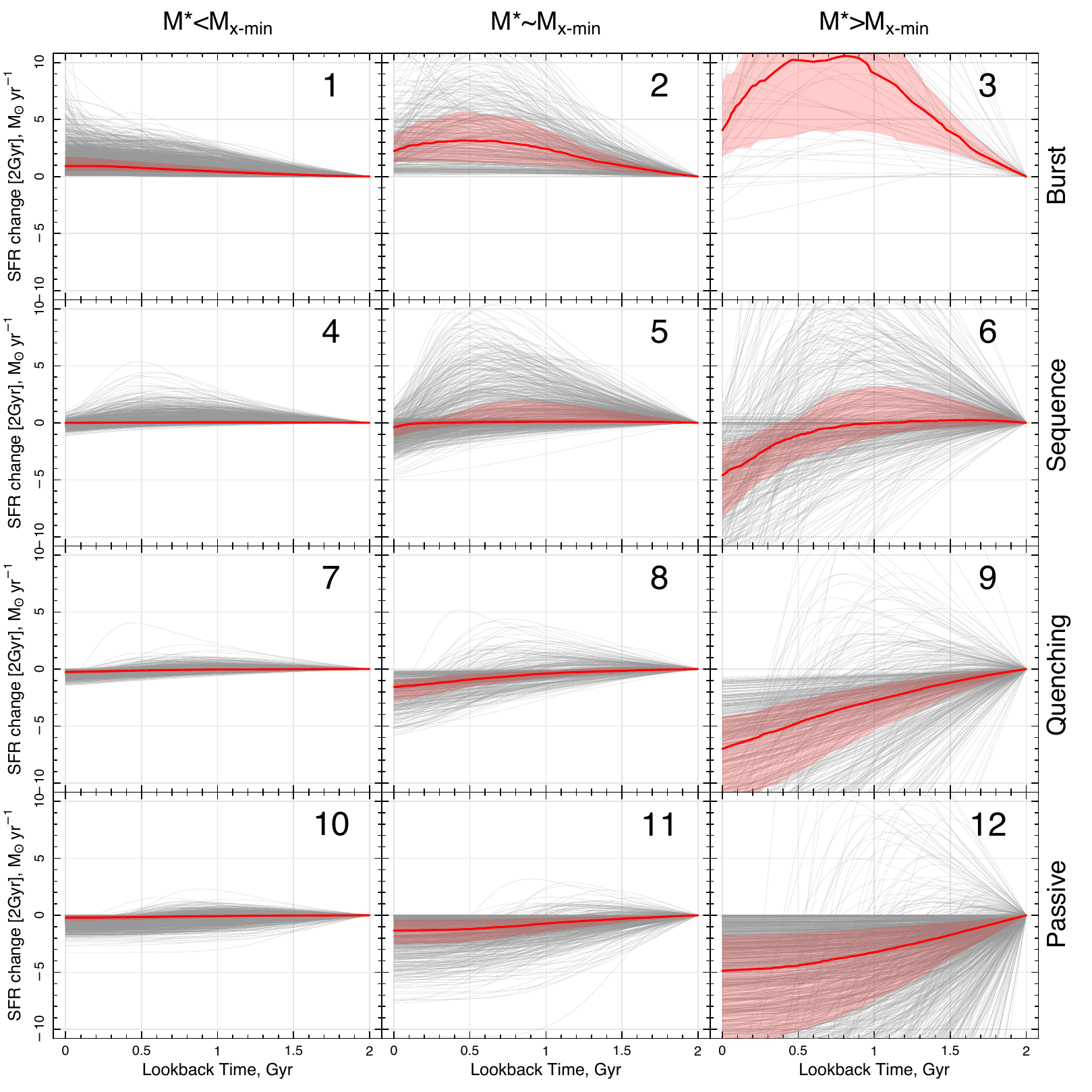}

\caption{The change in recent (last 2\,Gyr) SFH of galaxies in each of the 12 regions defined in Figure \ref{fig:selection} for galaxies at $0.4<z<0.55$, where look back time is relative to the galaxy's observation epoch. SFHs are scaled to the same point at 2\,Gyr lookback time and using a linear y-axis. While this loses information regarding the absolute normalisation of the SFH (removing the dynamic range), it allows for a more detailed analysis and comparison of the variation in SFH shape over the last 2\,Gyr in the galaxy's lifetime. }
\label{fig:z3_rel}
\end{center}
\end{figure*}

\section{Analysis}

\subsection{SFH variations across the sSFR-M$_{\star}$ plane}
\label{sec:var}

Within D22 we parameterised the shape of the $\sigma_{SFR}$-M$_{\star}$ relation, its evolution and the minimum SFR dispersion point (M$^{*}_{\sigma-min}$). In this series of papers we wish to explore the variation of SFHs across the plane and the astrophysical mechanisms that lead to the observed trends seen in this previous work. D22 argues that M$^{*}_{\sigma-min}$(z) represents a key parameter in the evolution of galaxies and defines the boundary between different evolutionary mechanisms that are shaping the sSFR-M$_{\star}$ plane. If this is true, we may expect to see different SFHs for galaxies in regions selected with respect to the M$^{*}_{\sigma-min}(z)$ point. However, stellar mass is likely not the only driver of variations in SFH, as recent star-formation activity itself will also significantly impact a galaxy's SFH. As such, we also explore SFH variations as a function of current sSFR. Traditionally, it has been assumed that a galaxy's position relative to the SFS is indicative of its recent \textcolor{black}{changes to star-formation}, with galaxies lying above the SFS at a given stellar mass being deemed `star-bursting', galaxies on the SFS having constant SFHs, and galaxies sitting below the SFS going through quenching \textcolor{black}{\citep[$e.g.$][]{Elbaz11, Schawinski14,Bremer18,Phillipps19, Salim23}}. To test this, we will first select regions of the sSFR-M$_{\star}$ plane based on stellar mass and position relative to the SFS, and explore the SFHs of galaxies in each sub-region. While the exact boundaries used in these sub-regions will subtly impact any result derived from this analysis, we opt to use reasonable selection regions (consistent with many previous works) and discuss new selection regions later in this paper.

In Figure \ref{fig:selection} we first select galaxies in different regions of the sSFR-M$_{\star}$ plane based on the M$^{*}_{\sigma-min}(z)$ point and the SFS. We define three stellar mass ranges, which evolve with redshift with respect to M$^{*}_{\sigma-min}(z)$ (gold vertical dashed lines in Figure \ref{fig:selection}), These represent galaxies at below the M$^{*}_{\sigma-min}(z)$ point (hereon referred to as low stellar mass), at the M$^{*}_{\sigma-min}(z)$ point, and above the M$^{*}_{\sigma-min}(z)$ point (hereon referred to as high stellar mass). We then also define four sSFR ranges with respect to the SFS at each epoch. Here the SFS is defined by taking all galaxies at sSFR$>10^{-10.25}$yr$^{-1}$ and M$_{lim}$ $<$ M$_{\star}$ $<10^{10.5}$\,M${\odot}$, determining the running median in 0.1\,dex bins and fitting a linear regression model to the medians. The resultant fits are shown as the red lines in Figure \ref{fig:selection} and are consistent with the SFS fits of \cite{Thorne21}. We note here that while the SFS may be a transitory feature of galaxies at a given epoch and that the position of a galaxy relative to the SFS may not be particularly informative in terms of its evolution (see later discussion), it does likely represent `typical' star-forming galaxies at a given stellar mass and epoch, and hence is an appropriate reference point. The cross-section of these stellar mass and sSFR cuts lead to 12 sub-regions in the sSFR-M$_{\star}$ plane shown in Figure \ref{fig:selection} at each epoch. We number each of these sub-regions 1-12 for ease of description, where the full region definitions are given in Appendix Tables \ref{tab:def} and \ref{tab:subsamp}. For ease of description, we define our sSFR cuts as `bursting', `sequence', `quenching' and `passive', from high to low sSFRs respectively. We note here that we are not suggesting all galaxies in these regions are bursting/sequence/quenching/passive galaxies, but simply use these terms for ease of describing our various sub-regions and following similar naming conventions to previous works.

\vspace{2mm}

We then explore the variation in \textsc{ProSpect}-derived SFHs for each region and epoch independently.  First, we take the parameterisation of the SFH for each galaxy from \cite{Thorne22} derive the SFHs over the last 2\,Gyr. \textcolor{black}{The choice of timescale with which to display these SFHs is largely subjective, and is somewhat determined based on the visual trends in the data (and we do not use the change over 2\,Gyr as a quantitative metric in our analysis). However, this also matches to the typical quenching timescale for galaxies \citep[$e.g.$][]{Bremer18,Phillipps19,Bravo23}. As in paper II of this series we will explore the physical driver of these quenching processes, we opt to show our SFHs over this timescale here.}  However, comparing the SFHs for our sample directly does not allow for a detailed comparison of the variations in SFHs across the SFR-M$_{\star}$ plane. This is largely due to the fact that the absolute values of the SFRs span over 7 orders of magnitude, and therefore to directly compare them on a single figure requires using a log-scaling, thus losing any fine details. To overcome this and directly compare the \textit{recent change} in SFHs, we compare the SFHs to the same SFR at 2\,Gyr look-back time and use a linear y-axis. This comparison allows us to purely explore the $change$ in star-formation in the galaxy's recent history and directly visualise the spreads in SFHs. These scaled SFHs are shown in Figure \ref{fig:z3_rel}, for just the $0.4<z<0.55$ redshift range. \textcolor{black}{We also show the median and interquartile range SFH in each panel as the red lines and shaded regions respectively}. Other epochs show similar trends, and we will later display the evolution of the trends in SFH.

 \begin{figure*}
\begin{center}
\includegraphics[scale=0.63]{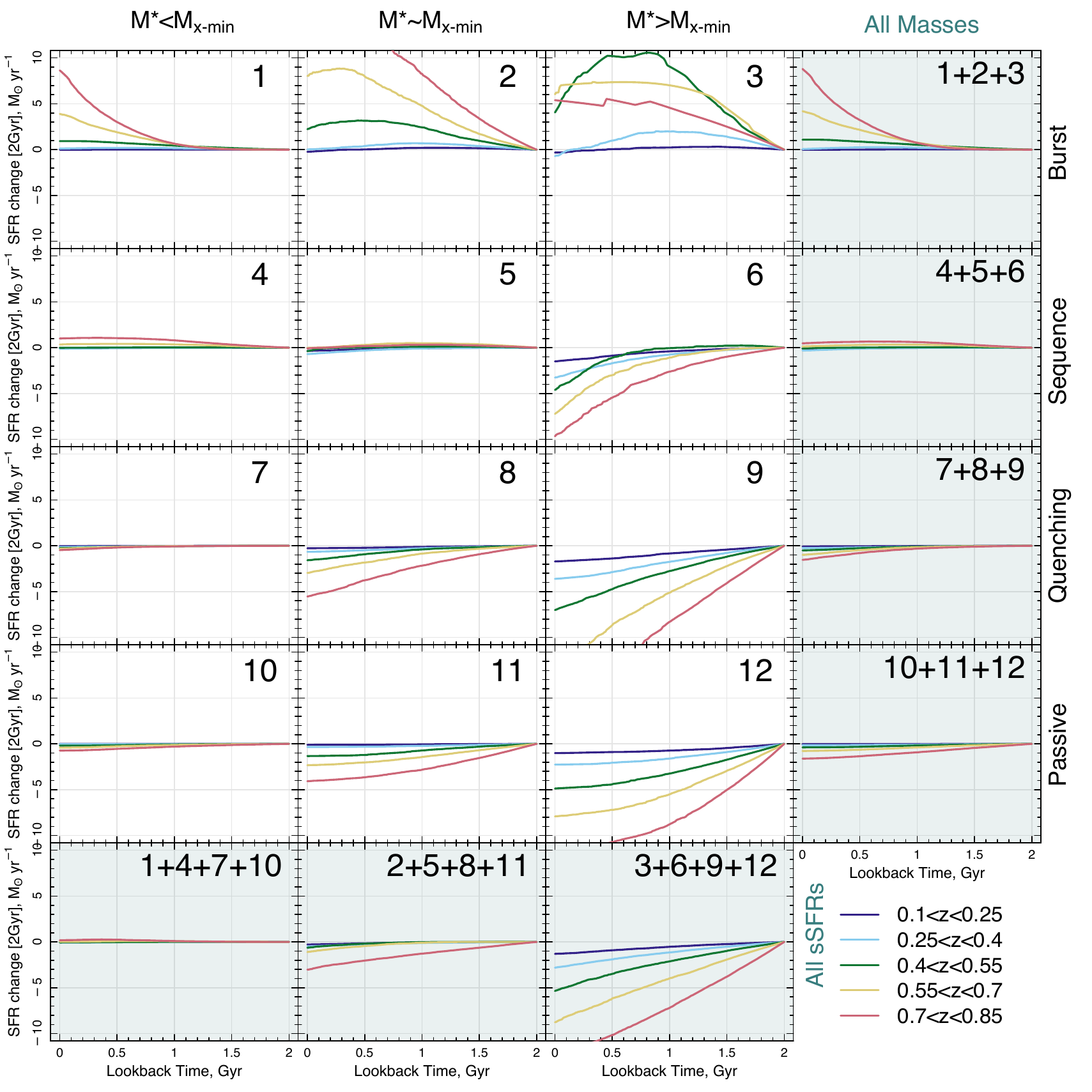}

\caption{Similar to Figure \ref{fig:z3_rel}, but showing the evolution of the median SFH change in each of our regions for the SFR-M$_{\star}$ relation. In addition, right-most column and bottom row show the distributions compressed in stellar mass and sSFR region respectively. }
\label{fig:zAll_rel}
\end{center}
\end{figure*}

\begin{figure*}
\begin{center}
\includegraphics[scale=0.63]{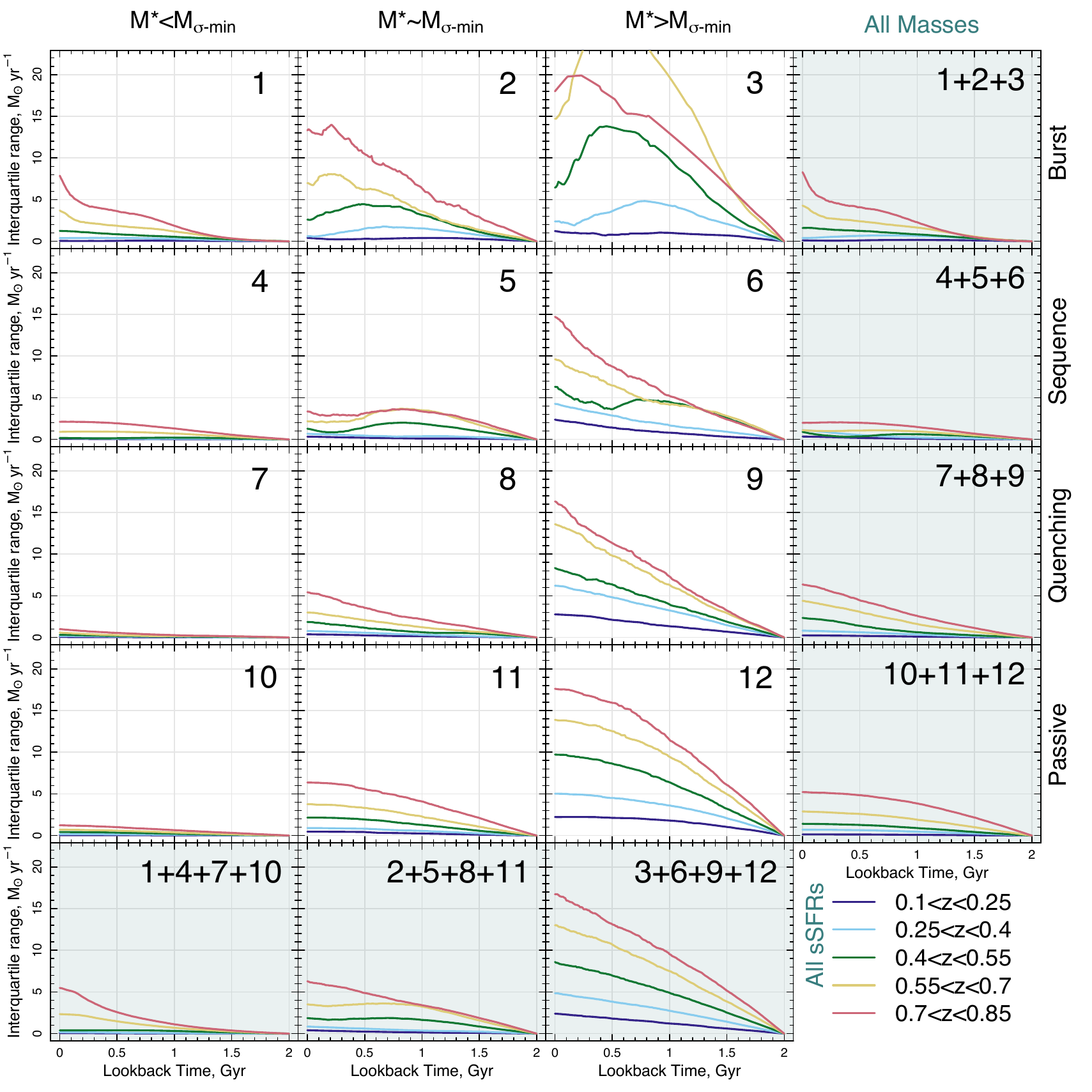}

\caption{The same as Figure \ref{fig:zAll_rel}, but showing the evolution of the interquartile range of SFH change in each of our regions for the SFR-M$_{\star}$ relation.}
\label{fig:zAll_rel_Qrange}
\end{center}
\end{figure*}

 \subsubsection{Choice of SFH metric}

In this work we opt to show the SFHs (and later use metrics to define recent change in SFH) in linear SFR-space, not log$_{10}$(SFR), and opt to use SFRs directly, over specific-SFRs. One could argue potential merits/pitfalls in using either change in log$_{10}$(SFR), or sSFR, or log$_{10}$(sSFR) as this metric instead, each of which providing different insights into the variation of SFHs and the movement of galaxies through this plane. For example, using a linear change in SFR one could argue that a change from 10\,M$_{\odot}$yr$^{-1}$ to 9\,M$_{\odot}$yr$^{-1}$ in comparison to a change of 2\,M$_{\odot}$yr$^{-1}$ to 1\,M$_{\odot}$yr$^{-1}$ should not be treated the same. However, likewise one could argue that using a log$_{10}$ change in SFR that a change from 100\,M$_{\odot}$yr$^{-1}$ to 10\,M$_{\odot}$yr$^{-1}$ in comparison to a change of 0.1\,M$_{\odot}$yr$^{-1}$ to 0.01\,M$_{\odot}$yr$^{-1}$ should not be treated the same. As such, the decision of which metric to use, is somewhat based on the science question at hand. In our analysis for this paper we have explored using each of these different metrics. However, we ultimately decide to use linear change in SFR here as we feel this best encapsulates what we are aiming to ultimately explore in this work - the astrophysical processes that are changing star-formation in galaxies, causing them to move through the sSFR-M$_{\star}$ plane and away from the SFS. $i.e.$ we wish to know in real terms how much the star-formation in galaxies has been affected, not relative to the current SFR or any other property. \textcolor{black}{To put this another way, what we ultimately wish to explore is how the stellar mass growth of galaxies is impacted by other factors, ultimately leading to processes such as quenching. As we are exploring the growth in stellar mass, we believe this is best encapsulated by change in linear SFR, which by definition is how a galaxy is growing in stellar mass over time, unscaled by any other factor. } 

\textcolor{black}{In this work we also present most of our results within the sSFR-M$_{\star}$ plane and/or split by stellar mass, allowing the reader to visually see variations as a function of stellar mass. Comparing galaxies across a broad range of stellar masses is problematic whichever metic is used, but displaying linear change in SFR provides the most directly relatable property with regard to processes such as, $e.g.$ quenching.  For example, one might argue that a linear change in SFR of 1\,M$_{\odot}$yr$^{-1}$ in two galaxies of vastly different stellar masses ($e.g.$ 10$^9$\,M$_{\odot}$ and 10$^{11}$\,M$_{\odot}$ ) should be treated differently. However, framing this in terms of stellar mass growth, we would argue the opposite, as both of these galaxies have had their stellar mass growth slowed by 1\,M$_{\odot}$yr$^{-1}$, irrespective of their absolute stellar mass. Yes, the \textit{impact} that this might have on the galaxy overall might be different for different stellar mass galaxies, but in terms of change in stellar mass growth, they are identical. We would argue that the effect of the stellar mass being slowed by 1\,M$_{\odot}$yr$^{-1}$ is directly what is happening to the galaxy, but the way this then impacts the galaxies is an interpretation, which can be affected by many things including the absolute stellar mass (but also the specifics of the underlying process that is causing the change to SFRs, the available gas supply, the morphology, the environment, etc).} 

\textcolor{black}{Further, looking at this in terms of quenching mechanisms (the scientific question we ultimately wish to probe), for example AGN or star-burst feedback either heating or driving out gas to suppress star-formation, if this is event is the same in both galaxies ($i.e.$ same energy output into the gas from an AGN or star-burst) it may affect the linear SFR more directly than the log$_{10}$(SFR). This is modulo the available gas supply, environment and/or gravitational potentail etc, but these are not directly correlated with stellar mass either. In summary, we wish to explore processes that impact stellar mass growth, and these are more directly related to linear SFR . Further interpretation is required to explore how this impacts galaxies over a range of other properties (including stellar mass). Here, we also aim to provide the reader with enough information to allow this interpretation for themselves. As such, in addition to the linear SFR changes showing in the main body of this paper, in Appendix \ref{app:log10} we do show versions of Figure \ref{fig:z3_rel} (and others) but using log$_{10}$(SFR), sSFR, and log$_{10}$(sSFR). This allows the reader to explore the variation in SFHs using whichever metric they prefer, and where appropriate, we comment on the difference between these metrics when quantifying our results.}

 \subsubsection{Trends in SFH variation}

\textcolor{black}{With our  choice of SFH metrics define and caveats discussed}, in Figure \ref{fig:z3_rel} (and somewhat in Figure \ref{fig:log_SFHs_z3Rel}) we see differences between the shape of the SFHs for galaxies above and below M$^{*}_{\sigma-min}(z)$. Low stellar mass systems at this epoch have relatively flat SFHs over their last 2\,Gyrs at all sSFRs relative to the SFS. So while there is a large variation in the absolute values of their sSFRs ($i.e.$ Figures \ref{fig:selection}), the actual shape of the their SFHs are very similar and shows little change over the last 2\,Gyr. The exception to this is a fraction of the `burst' population (region 1), which show some rapidly increasing star-formation rates  in a galaxy's recent history. We note again that while the \textsc{ProSpect} SED analysis is not sensitive to very short-timescale variations in SFH, increases in SFRs on the timescale of those seen in region 1 are detectable, and as such, the lack of these in other panels is not simply a consequence of the fitting procedure and data limitations. 

In high stellar mass galaxies, we see much greater variation in recent SFHs when subregions are selected relative to the SFS. In the `burst' region (3), we find galaxies that have recently undergone a rise in their SFRs, but are now rapidly on the decline ($i.e.$ they lie above the SFS but are in fact declining in star-formation at their observation epoch), in the `sequence' region (6) we see a mixed bag of different SFHs, but they are also generally on the decline, in the `quenching' region (9) we see systems that are predominantly declining in star formation, while in the `passive' region (12) we predominantly see galaxies that declined in star formation in the recent past, but are now relatively constant ($i.e.$ they quenched in the past and are now passive - as expected). As such, once again this suggests that selecting galaxies relative to the SFS, may not be appropriate to identify galaxies with common SFHs, as some regions show a broad variety of SFH shapes.   

At the M$^{*}_{\sigma-min}(z)$ point we essentially see a mid-way mixture between the low- and high-stellar mass cases, this is to be expected as it is potentially the dividing region between two different regimes of SFR evolution. Interestingly, this also suggests that, at this epoch at least, the high stellar mass end of the SFR-M$_{\star}$ plane is evolving in terms of star-formation and stellar mass growth very differently to the low stellar mass end. For example, galaxies at the low stellar mass end are evolving self-similarly with constant SFHs (modulo the burst population), while at the high stellar mass end galaxies are predominantly declining in SFR. This likely results in the observed turnover in the SFS \citep[$e.g.$][]{Thorne21} and is consistent with the evolution of the SFS shown in Figure 10 of \cite{Davies22}. The effect this has on the evolution of points within the overall SFR-M$_{\star}$ plane and the SFS will be discussed in later sections.

\vspace{5mm}

It is next interesting to consider how this variation in SFH across the sSFR-M$_{\star}$ plane evolves with redshift. While we only display Figure \ref{fig:z3_rel} at a single epoch, we next compile the median SFH change in each subregion and at each epoch in Figure \ref{fig:zAll_rel}. However, just displaying the median SFH does not tell the whole story as this does not parametrise the \textit{spread} in SFHs. As such, we also display the evolution of the interquartile range of SFHs in Figure \ref{fig:zAll_rel_Qrange}. \textcolor{black}{In both of these figures we also show the median SFH and interquartile range of SFHs for all stellar masses combined in a given sSFR region (rightmost column), and for all sSFR regions combined at a given stellar mass range (bottom row).}     

 In combination, these Figures display a number of interesting trends:\\

\noindent $\bullet$ Firstly, at low stellar masses (panels 1,4,7,10), galaxies show very little overall change in their \textcolor{black}{$linear$} SFHs over that last 2\,Gyr and very little spread in SFH shape across all epochs (both the median SFH change is very close to zero, and the interquartile range is small at all epochs). The only exception to this is the `burst' population (region 1), which shows rapidly increasing SFRs. In combination with the large spread in t=0 sSFRs in the low stellar mass population ($i.e.$ in Figure \ref{fig:selection}), this suggests that low-mass galaxies have a broad range of \textit{normalisation} in their SFH, but little variation in slope/shape when not in the 'burst' region. $i.e.$ they are all forming stars at a constant rate, but that rate is dependent on a quantity other than just stellar mass. This is intriguing and warrants further study, and will be explored in later sections and paper II. However, to first order it appears that the large SFR dispersion seen at low stellar masses, and by extension the large $\sigma_{SFR}$, \textcolor{black}{may not} $only$ be caused by stochastic bursting/quenching events (as previous works have suggested), but also by a large spread in longer-duration SFH normalisation. \textcolor{black}{$i.e$ bursty galaxies above the SFS at low stellar masses contribute to the observed scatter, but also lower sSFR galaxies below the SFS with a broad normalisation range of relatively constant SFHs also add to this scatter. Looking at this a different way, if the large scatter in sSFRs below the SFS at low stellar masses was caused solely by SFR stochasticity, we may expect to see galaxies below the SFS which are rapidly increasing in star-formation to return to a self-regulated state, and we do not. One potential caveat here is that short timescale variations in star-formation may be washed out by the \textsc{ProSpect} analysis, which fits a smooth log-normal distribution. However, as discussed previously, due to the ability of \textsc{ProSpect} to fit a negative time peak to the SFH, systems with a recent t=0 burst, and no older stellar populations, will be captured in our analysis and should be evident here - these are the rapidly increasing systems in panel 1. \textcolor{black}{It is also possible to directly test the ability of \textsc{ProSpect} to recover bursty-like SFHs by fitting simulated galaxies and exploring the successes of recovering the recent SFH \citep[$e.g.$][]{Bravo22, Bravo23, Lagos24}. This is undertaken in Appendix \ref{sec:shark}, where we find no systematic bias in \textsc{ProSpect}'s ability to recover the slope of the recent SFH in simulated bursty, low stellar mass galaxies. }   Its is also worth highlighting that this statement may only true when considering $linear$ SFHs. If we instead consider the log$_{10}$(SFHs) shown in Figure \ref{fig:log_SFHs_z3Rel}, we do see larger diversity in SFHs. First, this indicates that the choice of how the SFHs are presented can significantly impact the inferences derived. Secondly, while there is a diversity of log$_{10}$(SFHs) at the low stellar mass end, except for in the `burst' region (1) this does not appear to be significantly different (except in terms of numbers) from the diversity in the other stellar mass ranges (i.e. across rows in Figure \ref{fig:log_SFHs_z3Rel}). If the scatter below the SFS at low stellar masses is purely driven by stochasticity, we would likely expect there to be significant differences in the distribution of SFHs in comparison to higher stellar mass samples.} \\     

\noindent $\bullet$ At the minimum SFR dispersion point (M$^{*}_{\sigma-min}$ - panels 2,5,8,11) the median SFHs of galaxies show trends as expected if SFH is directly related to the current sSFR relative to the SFS. $i.e.$ that in the `burst' population galaxies have been bursting, in the `sequence' population galaxies show very flat and consistent SFHs, in the `quenching' population galaxies are declining in SFR, and in the `passive' population galaxies have declined in star formation at some point over the last 2Gyr, but are now tending to constant SFRs. However, we do find a relatively large spread in SFH shapes (as traced by the interquartile range) in all populations, suggesting that at this stellar mass, we do find large variety of galaxy types. \\

\noindent $\bullet$ At high stellar masses (panels 3,6,9,12) we find that the majority of galaxies are declining in star formation at all sSFRs, even in the `burst' region (except for the very highest redshift). The most rapidly declining galaxies are in the quenching region, but we do also see strongly declining SFRs in the region that would typically be defined at the SFS. We also find that this population has the largest variation in SFH shapes (as traced by the interquartile range). This is also clear in Figure \ref{fig:z3_rel}, where we see that the high stellar mass population does in fact contain galaxies that have a diverse range of different SFH shapes. Galaxies have been through recent increases in star formation and then a rapid decline, gradual declines in star formation and relatively constant star formation (either as star-forming, region 6, or passive systems, region 12), largely irrespective of their position relative to the SFS. $i.e.$ at these stellar masses, even a galaxy that is on the SFS and has sSFR$\sim10^{-9.5}$\,yr$^{-1}$ (which would be classed as star forming based on the sSFR-M$_{\star}$ plane) can have rapidly declining SFRs over the last $\sim$Gyr. \\

  \noindent $\bullet$ There are strong trends with stellar mass across our sample (bottom row of figures), with low stellar mass galaxies being dominated by constant (but star-forming) or increasing SFHs, and high-stellar mass galaxies being dominated by either constant (but passive) or rapidly declining SFHs. This indicates that stellar mass must play a large role in determining the recent SFH of a galaxy (as expected). However, these trends are not constant with epoch or sSFR, such that stellar mass can not be the only factor at play. \\

 \noindent $\bullet$ Across the sample we find that as we move from the high to low redshift Universe, galaxies at all stellar masses and sSFRs are tending towards more constant and stable SFHs. The extremes of both rapidly increasing and rapidly declining SFRs, and the variation in SFHs is reducing with time as the Universe tends to a more stable and constant state. This is consistent with downsizing, the overall decline in star-formation activity in the Universe and a decrease in galaxy major mergers at these stellar masses (which can cause dramatic changes to SFHs).  \\

In combination, these results show that our traditional picture of being able to select galaxies which are bursting, constant, quenching or passive based on their position relative to the SFS may be flawed as this statement does not hold true at all stellar masses. For example, galaxies that sit just below the SFS at low mass (which may have been thought to be in the process of quenching) show very stable and constant SFHs, while galaxies that sit on or above the SFS at high stellar masses (which may have been thought to be star-forming or bursting) actually show a rapid decline in their recent SFH. \textcolor{black}{For the high stellar mass end,} this is very clear when exploring linear SFHs, but also to a lesser extent when using the other types of SFH shown in Appendix \ref{app:log10}. The trends are weaker, but for example, high stellar mass galaxies on the SFS are declining in their recent SFH, both in linear and log$_{10}$(SFH) ($e.g.$ in panel 6 of Figure \ref{fig:log_SFHs_z3Rel}). \textcolor{black}{However, we also indicate that in terms of the low stellar mass end, linear and log$_{10}$ SFHs do show distinctly different trends.} To explore this further we will next define a single metric to capture the galaxy's recent change in SFH, which aims to better parametrise how a galaxy is moving through the SFR-M$_{\star}$ plane, and use this metric to better define regions of the SFR-M$_{\star}$ plane that share common SFHs.

\subsection{The $\Delta$SFH$_{200\mathrm{\,Myr}}$ of galaxies}
\label{sec:deltSFR}

In order to provide this simple metric for how galaxies are currently moving through the sSFR-M$_{\star}$ plane we calculate the change in star formation over the last 200\,Myr as:

\begin{equation}
\Delta$SFH$_{200\mathrm{\,Myr}}=SFR(t_{LB}=0)-SFR(t_{LB}=200Myr)
\label{eq:slope}
\end{equation}

\noindent where $SFR(t_{LB})$ is taken from the galaxy's \textsc{ProSpect}-derived SFH. This essentially reduces the shape of the recent SFH to a single value. We opt to use 200\,Myr in this analysis as the slope of the SFHs of galaxies are relatively constant over this timescale and this is likely close to the shortest timescale over which \textsc{ProSpect} can robustly measure SFR changes for all galaxy types. For example, over the 2\,Gyr timescale shown in, $e.g.$, Figure \ref{fig:z3_rel} there the many galaxies that show an initial raise then then decline in SFR. Such systems would not have their recent SFH accurately represented with the above metric, but using a longer timescale. Most galaxies do have a relatively constant slope over the last $\sim$1\,Gyr in their \textsc{ProSpect} SFH. However, we opt for a shorter timescale than $\sim$1\,Gyr as (especially at the high redshift end), many galaxies are younger than 1\,Gyr. Thus, measuring their $\Delta$SFH over this duration would not produce a robust measure of their SFR change. $i.e.$ for these sources SFR$(t_{LB}=1Gyr)=0$ and thus $\Delta$ SFH$_{1Gyr}$=SFR$(t_{LB}=0)$ and our SFH change is just measured as the current SFR.  However, for completeness, we do repeat all of the subsequent analysis in this paper using \textcolor{black}{$\Delta$SFH$_{100\mathrm{\,Myr}}$} and $\Delta$SFH$_{1\mathrm{\,Gyr}}$  time steps and find that it does not significantly affect the results. \textcolor{black}{For completeness, in Appendix \ref{app:log10}, we  show how the overall distribution of $\Delta$SFH values changes when using $\Delta$SFH$_{100\mathrm{\,Myr}}$ (Figure \ref{fig:slopeRanges100}), highlighting that the overall trends remain the same. }

\textcolor{black}{As noted previously, in our analysis we opt to use linear change in SFR as our metric for the recent change in SFH.  However, in Appendix \ref{app:log10} we once again show how this choice impacts the overall distribution of $\Delta$SFH values across the sSFR-M$_{\star}$ plane (Figure \ref{fig:slopeRanges}) and discuss similarities/differences compared to considering log change in SFR. While this is discussed in Appendix \ref{app:log10}, we note that the overall trends are somewhat similar when using log$_{10}$(SFR), and the choice of new selection regions for common SFHs (described in the next section) largely hold true. }       

Finally, to test the validity of using $\Delta$SFH$_{200\mathrm{\,Myr}}$ from \textsc{ProSpect}-derived SFHs to define recent changes in the true SFH of galaxies, in Appendix \ref{sec:shark} we use the \textsc{Shark} \citep{Lagos18} semi-analytic model to compare $\Delta$SFH$_{200\mathrm{\,Myr}}$ values from \textsc{ProSpect} fits to simulated galaxies with the true SFH. In summary, Appendix \ref{sec:shark} suggests that while there are differences, the \textsc{ProSpect} SFHs (and the derived $\Delta$SFH$_{200\mathrm{\,Myr}}$ values), are adequate to explore the overall evolution of galaxies in subregions of the sSFR-M$_{\star}$ plane, but are likely not sensitive to short time-scale fluctuations in SFH for individual sources; as previously discussed. \textcolor{black}{This is also reported in \cite{Bravo22} when directly exploring the \textsc{ProSpect} fits to \textsc{shark} galaxies.}

\vspace{2mm}

\begin{figure*}
\begin{center}
\includegraphics[scale=0.55]{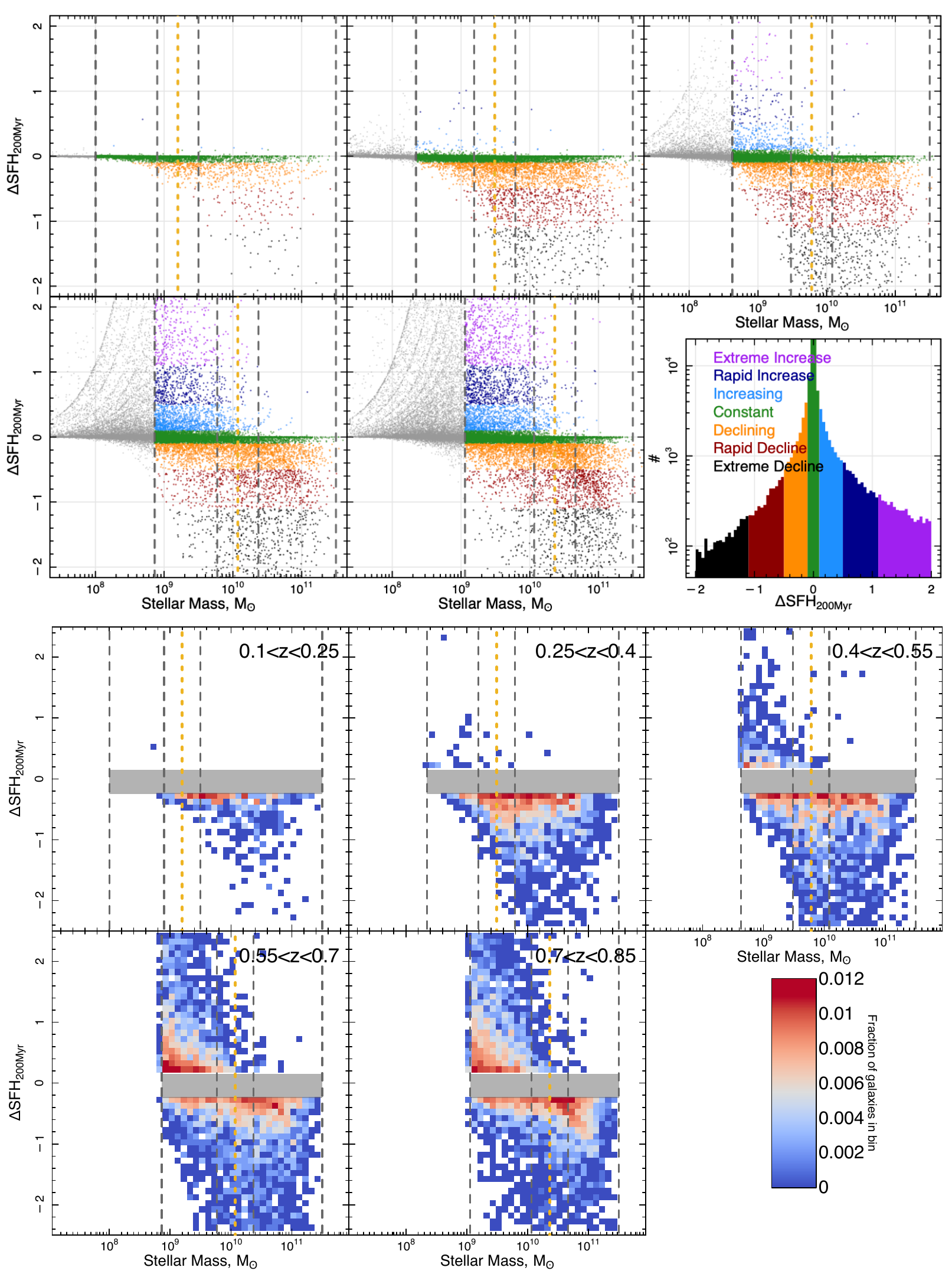}

\caption{The variation of SFR over the last 200\,Myr ($\Delta$SFH$_{200\mathrm{\,Myr}}$) against stellar mass. In the top panels we show the raw data, where points above our stellar mass completeness limit at each epoch are coloured in bands to aid the eye. In the bottom panels we show the number density of galaxies in this plane with the bins around $\Delta$SFH$_{200\mathrm{\,Myr}}$=0 and below the stellar mass selection limit masked to highlight other features in the distribution. As in other panels the dashed vertical gold line shows the evolution of M$^{*}_{\sigma-min}$. The M$^{*}_{\sigma-min}$ point appears to trace a dividing line between sources which are predominantly increasing in star formation and those which are declining. This is more extreme on the increasing side, where we essentially see very few galaxies above M$^{*}_{\sigma-min}$ which are increasing in SFR. }
\label{fig:M_sSFRslopeGyr}
\end{center}
\end{figure*}

With caveats for choice of SFH metric aside, in Figure \ref{fig:M_sSFRslopeGyr}, we show the variation in $\Delta$SFH$_{200\mathrm{\,Myr}}$ as a function of stellar mass at each epoch. The top panels display the data points colour-coded by $\Delta$SFH$_{200\mathrm{\,Myr}}$ ranges as shown in the bottom right panel. We split the $\Delta$SFH$_{200\mathrm{\,Myr}}$ values into seven different ranges: \\

\noindent Extreme declining: $\Delta$SFH$_{200\mathrm{\,Myr}}<=-1.1$M$_{\odot}$yr$^{-1}$  \\

\noindent Rapid declining: $-1.1<\Delta$SFH$_{200\mathrm{\,Myr}}<=-0.5$M$_{\odot}$yr$^{-1}$ \\

\noindent Declining: $-0.5<\Delta$SFH$_{200\mathrm{\,Myr}}<=-0.1$M$_{\odot}$yr$^{-1}$ \\

\noindent Constant: $-0.1<\Delta$SFH$_{200\mathrm{\,Myr}}<=0.1$M$_{\odot}$yr$^{-1}$ \\

\noindent Increasing: $0.1<\Delta$SFH$_{200\mathrm{\,Myr}}<=0.5$M$_{\odot}$yr$^{-1}$ \\

\noindent Rapid increasing: $0.5<\Delta$SFH$_{200\mathrm{\,Myr}}<=1.1$M$_{\odot}$yr$^{-1}$ \\

\noindent Extreme increasing: $\Delta$SFH$_{200\mathrm{\,Myr}}>1.1$M$_{\odot}$yr$^{-1}$ \\

\noindent This colour-coding is simply defined by eye and will aid in the description of later figures to highlight sources that have common $\Delta$SFH$_{200\mathrm{\,Myr}}$ values (and therefore common recent SFHs). The bottom panels display the number density of points in bins of $\delta$(log$_{10}$(M$^*$))=0.1 and $\delta(\Delta$SFH$_{200\mathrm{\,Myr}}$)=0.1, with the $\Delta$SFH$_{200\mathrm{\,Myr}}\sim0$ bins masked to increase the dynamic range of features in the increasing/declining populations. We also note here that there is a strong ridge line at $\Delta$SFH$_{200\mathrm{\,Myr}}=0$ in the top panels (constant SFR), which is not particularly clear. However, these figure are largely designed to show the extremes of $\Delta$SFH$_{200\mathrm{\,Myr}}$.

Considering these Figures, firstly it is evident that at all epochs that there are strong trends in the extreme values of $\Delta$SFH$_{200\mathrm{\,Myr}}$ with stellar mass. We see that almost all of the population that has positive $\Delta$SFH$_{200\mathrm{\,Myr}}$ values (increasing in star-formation) occur at low stellar masses. We do see galaxies with negative $\Delta$SFH$_{200\mathrm{\,Myr}}$ values (decreasing in star-formation) at all stellar masses, but care must be taken here as there is a sharp boundary for negative $\Delta$SFH$_{200\mathrm{\,Myr}}$ values below M$^{*}<10^{10}$M$_{\odot}$ - where we do not see low-stellar mass galaxies with declining SFHs. This may be a consequence of the fitting methodology/selection limit. However, this does not account for the paucity of positive $\Delta$SFH$_{200\mathrm{\,Myr}}$  sources at the high mass end. In fact, the stellar mass point at which we stop seeing positive $\Delta$SFH$_{200\mathrm{\,Myr}}$ sources, does qualitatively appear to trace the M$^{*}_{\sigma-min}$ point at all epochs (vertical gold dashed line), which is intriguing. \textcolor{black}{We also highlight that the banded regions at high $\Delta$SFH$_{200\mathrm{\,Myr}}$  and low stellar masses are a consequence of the lack of very low metallicity stellar population templates in the \textsc{ProSpect} fitting which then hit a boundary of the available parameter, but that all of these sources sit below our stellar mass selection limits - so are not used in our analysis.}

Considering the density of galaxies across this plane, we also see an interesting feature in the distribution of points. There appears to be a faint ridge line of high-density in the sources with negative $\Delta$SFH$_{200\mathrm{\,Myr}}$ (declining in star formation), at $\sim10^{10.75}$M$_{\odot}$ (most markedly in the highest redshift panel). \textcolor{black}{This is also seen in similar figure using the other SFH types ($i.e.$ as in Appendix \ref{app:log10})}. While this is potentially an artefact of the fitting process, it is intriguing that there may be a specific stellar mass at which galaxies are more-likely displaying decreasing SFHs. This is close to M$^{\star}$ (characteristic stellar mass) and coincides with a decline in the number of galaxies with constant $\Delta$SFH$_{200\mathrm{\,Myr}}$ at higher stellar masses. It is unclear how this could be produced by our fitting process or a selection effect, and could indicate a quenching mechanism that occurs at a specific stellar mass, driving galaxies away from constant SFR and into the declining region of this parameter space. \textcolor{black}{This is potentially also seen in the works of \cite{Iyer19} and \cite{ Leja22}, who find an increased number of declining SFHs at these stellar masses.}  This will be discussed further in paper II.

\begin{figure*}
\begin{center}
\includegraphics[scale=0.68]{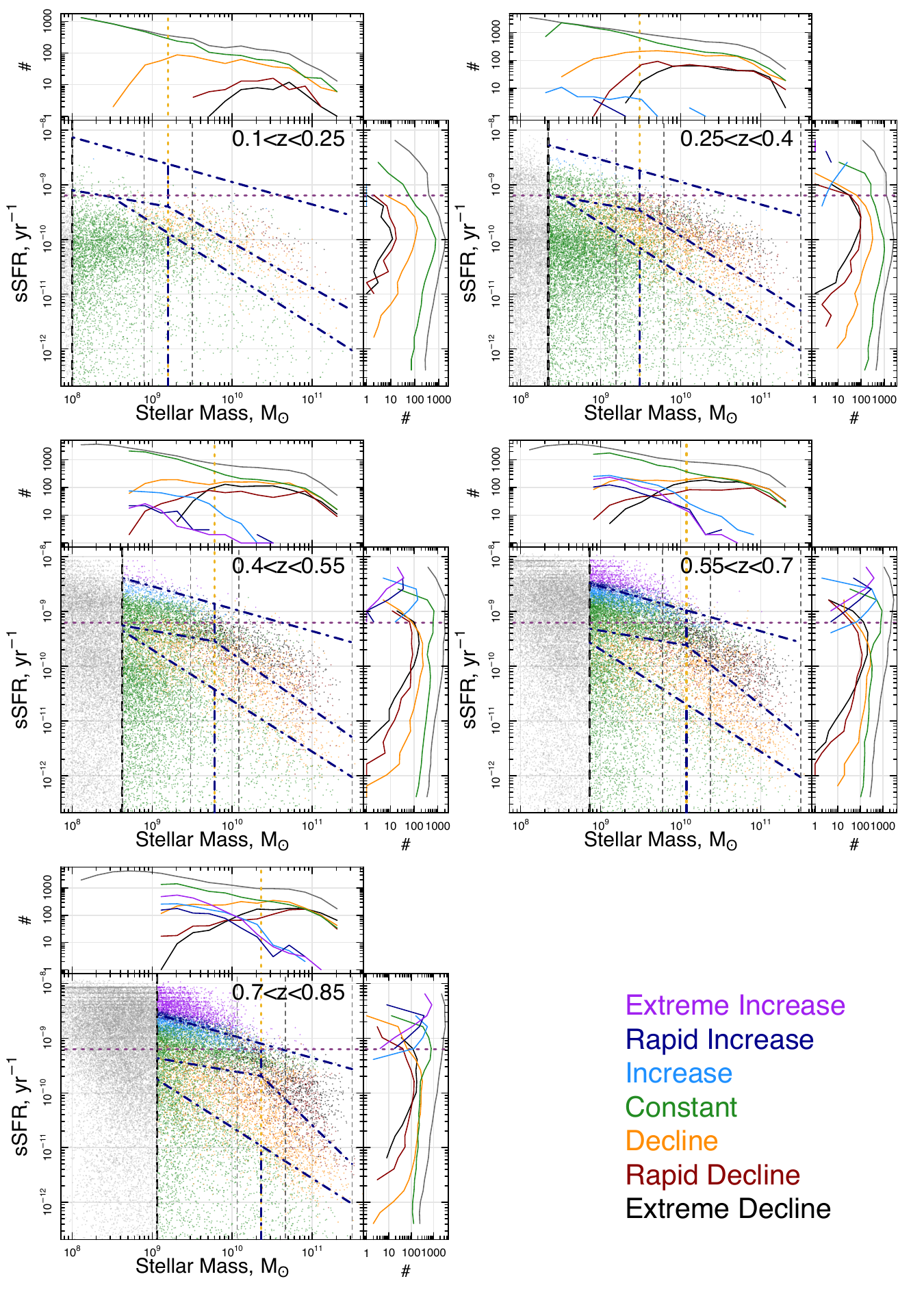}

\vspace{-2mm}

\caption{The variation of SFR over the last 200\,Myr ($\Delta$SFH$_{200\mathrm{\,Myr}}$) in relation to position in the sSFR-M$_{\star}$  plane. Points are colour-coded based on bands in $\Delta$SFH$_{200\mathrm{\,Myr}}$ described in Figure \ref{fig:M_sSFRslopeGyr}. Each panel shows a different epoch with the sSFR-M$_{\star}$  plane  displayed and histograms of the distributions compressed along each axis. As in other figures the dashed vertical gold line shows M$^{*}_{\sigma-min}$ at each epoch, the dashed grey vertical lines display our selection boundaries. In this figure we also plot a horizontal dark purple line at sSFR=$10^{-9.2}$\,yr$^{-1}$, which appears to roughly divide the SFH increasing and declining populations at all epochs. We also define a number of new regions to select galaxies with common recent SFHs (see text and Figure \ref{fig:commonBins} for details), which are displayed as navy dot-dashed line.}
\label{fig:slope}
\end{center}
\end{figure*}

\subsubsection{Variation of $\Delta$SFH$_{200\mathrm{\,Myr}}$ values across the sSFR-M$_{\star}$ plane}

\vspace{2mm}

Taking this one stage further, we next display the distribution of $\Delta$SFH$_{200\mathrm{\,Myr}}$ values across the full sSFR-M$_{\star}$ plane in Figure \ref{fig:slope}. We clearly see that galaxies with different $\Delta$SFH$_{200\mathrm{\,Myr}}$ values do populate different regions of this parameter space. For example, galaxies with rapidly increasing SFR (blue/purple points) all sit above the SFS at low stellar masses, the SFS is largely composed of galaxies with constant SFRs (green points) at below $\sim$M$^{*}_{\sigma-min}$, but is largely composed of sources with rapidly declining SFR (black/red points) above this. We also find that the stellar mass boundary between rapid/extreme declining population (black/red points) and constant star-formation (green points) along the SFS appears to evolve with M$^{*}_{\sigma-min}$, once again suggesting that it may represent a boundary point between different galaxy evolution processes. We also find that the upper stellar mass at which sources show rapidly increasing SFRs also appears to follow M$^{*}_{\sigma-min}$, potentially indicating that this population of galaxies with rapidly increasing SFRs that contribute to the large dispersions in SFRs at the low stellar mass end, and the SFR dispersion is a minimum where they no longer occur.  This Figure also displays that sources with constant/increasing SFRs occur at lower and lower stellar masses as the universe evolves, with more galaxies becoming constantly star-forming with time. We note that if we calculate the same $\Delta$SFH$_{200\mathrm{\,Myr}}$ metric for the \textsc{Prospect} GAMA \citep{Bellstedt20}  at $z<0.1$, only $\sim1$\% of the M$_{\star}>10^{10.5}$M$_{\odot}$ population is increasing in star-formation. As such, this is not a quirk of the DEVILS sample (albeit both analyses used the same \textsc{ProSpect} methodology).  These results are consistent with the picture of downsizing (as star-formation occurs in lower stellar mass systems with time), but is also very consistent with the results of D22, which suggest that the M$^{*}_{\sigma-min}$ point moves to lower stellar masses as the Universe evolves because the high dispersion population occurs at lower stellar masses with time. Here, we see a similar picture, but traced by recent SFH, not SFR dispersion. 

Putting this another way, it appears that there is an upper stellar mass limit at which galaxies have either constant or increasing star-formation. Above this, they begin to decline in star-formation and drop off the SFS (leading to the turn-over in the SFS). This stellar mass point can be observationally traced by the minimum dispersion point along the SFS.   

We do also find a somewhat clear sSFR separation between sources with increasing and declining SFHs at around sSFR=$10^{-9.2}$\,yr$^{-1}$ . We display this as the dark purple horizontal dotted line in Figure \ref{fig:slope}. Interestingly the intersection of this line and the gold vertical M$^{*}_{\sigma-min}$ line, appears to trace the upper boundary of the SFS at all epochs, despite not being defined by the SFS at all. We note here, that D22 does suggest a fixed sSFR point above which galaxies can have bursty star-formation leading to large dispersion in sSFRs, and that this point potentially occurs at sSFR=$10^{-9.6}$\,yr$^{-1}$  (close to the value found here).

\begin{figure*}
\begin{center}
\includegraphics[scale=0.62]{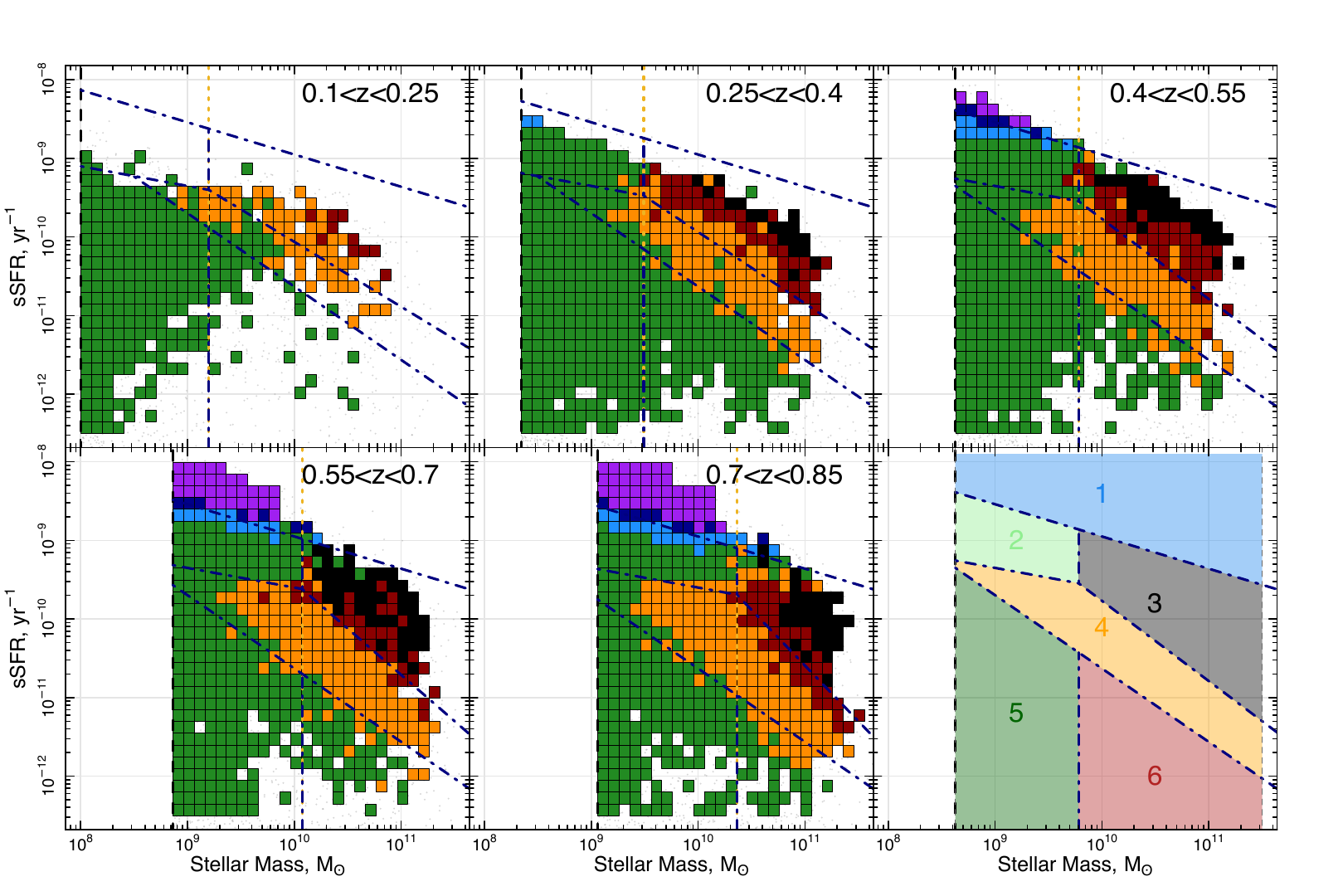}

\vspace{-2mm}

\caption{The variation of $\Delta$SFH$_{200\mathrm{\,Myr}}$ values across the SFR-M$_{\star}$ plane. At each epoch we select $\Delta$log$_{10}$(M$_{\star}$)=0.1 and $\Delta$log$_{10}$(sSFR)=0.15 bins above our stellar mass limit. In each bin we determine the median $\Delta$SFH$_{200\mathrm{\,Myr}}$ value and colour-code by the $\Delta$SFH$_{200\mathrm{\,Myr}}$ banding in previous figures. Bin colours are only shown if $>$3 galaxies are found in a bin. The dashed gold vertical line displays the minimum SFR dispersion point at each epoch. Populations with different $\Delta$SFH$_{200\mathrm{\,Myr}}$ values occupy different regions of this parameter space. We then use these to select new regions that isolate common recent SFHs - dashed blue lines.  The bottom right panel of this figure displays our new selection regions (just showing the $0.4<z<0.55$ bin) defined as: 1) SF increasing (purple and blue), 2) SFS (green and high sSFR), 3) Rapid Quenching (black and red), 4) Slow quenching (orange), 5) Low mass passive? (green, low sSFR and low stellar mass), 6) Passive (green, low sSFR and high stellar mass).   }
\label{fig:commonBins}
\end{center}
\end{figure*}

\subsection{New selection regions for galaxies with common recent SFHs}
\label{sec:newreg}

Given that the populations colour-coded based on recent SFH fall into relatively distinct regions of the sSFR-M$_{\star}$ plane, we can now define new selection regions that aim to identify galaxies with common recent SFHs (and do not only use position relative to the SFS). The properties of galaxies in these regions and the evolution of these section boundaries will encode important information regarding the evolution of star-formation in galaxies, and elucidate the potential mechanisms that move galaxies through this plane. As such, we will first produce these new selection regions based on the distribution of points in Figure \ref{fig:slope} and, in paper II, explore the properties of sources in our new regions. 

To define these regions, we first look at the dominant $\Delta$SFH$_{200\mathrm{\,Myr}}$ value in each position of the sSFR-M$_{\star}$ plane. In Figure \ref{fig:commonBins} we split the plane above our stellar mass limits into $\Delta$log$_{10}$(M$_{\star}$)=0.1 bins and $\Delta$log$_{10}$(sSFR)=0.15 bins, and calculate the median $\Delta$SFH$_{200\mathrm{\,Myr}}$ value in each bin. We then colour-code the bin by the $\Delta$SFH$_{200\mathrm{\,Myr}}$ banding used previously. Here we see that there is clear separation between populations showing different $\Delta$SFH$_{200\mathrm{\,Myr}}$ values and that these populations evolve in position with time. This Figure essentially accentuates the trends seen in Figure \ref{fig:slope}, and clearly indicates the bounding between common SFHs.  We then use these populations to define subregions of the sSFR-M$_{\star}$ plane that contain galaxies with similar recent SFHs. These regions are described below.  For clarity our resultant regions are also shown in Figure \ref{fig:slope} and numbered in the bottom right panel of Figure \ref{fig:commonBins}. We will refer to these by number in our subsequent description.

\begin{figure*}
\begin{center}
\includegraphics[scale=0.13]{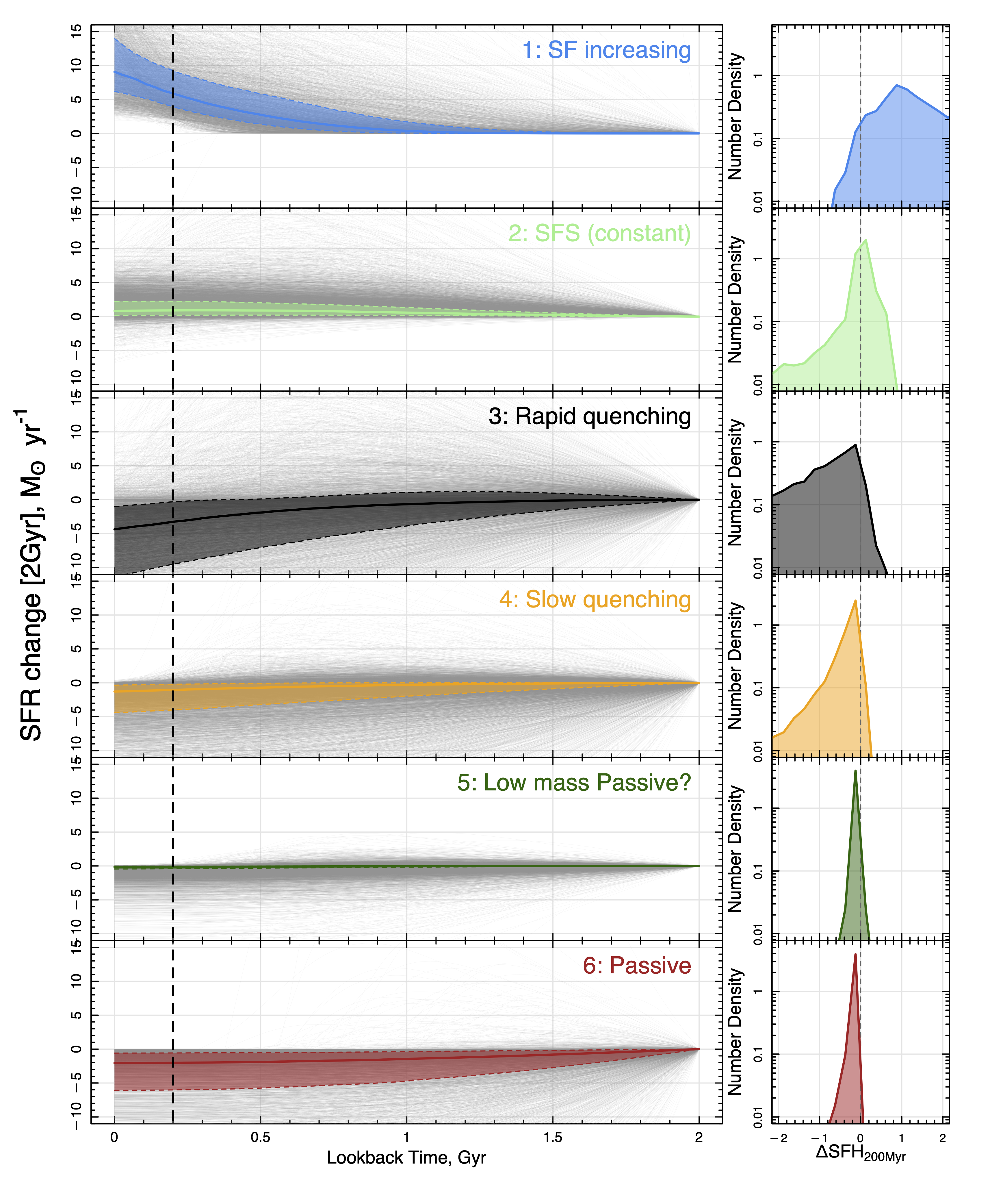}

\caption{Comparison of the SFH changes over the last 2\,Gyr (left) and $\Delta$SFH$_{200\mathrm{\,Myr}}$ distribution (right) for galaxies selected from the regions defined in Figure \ref{fig:slope}. In this figure all epochs are combined. In the left column, Individual SFHs are shown as faint grey lines, the coloured line displays the median SFH and coloured shaded region bounded by dashed line the interquartile range. The dashed vertical line displays 200\,Myr - the epoch over which the slope of the recent SFH is calculated in the right column. We show that our new selection regions do in fact select galaxies with common SFHs and both the SFH shape the distribution of $\Delta$SFH$_{200\mathrm{\,Myr}}$ values is consistent for the type of galaxy each region is aiming to select. $i.e.$ the SFS region selects SFHs with a flat slope and low scatter, while the `Rapid Quenching'  region selects declining SFHs with low $\Delta$SFH$_{200\mathrm{\,Myr}}$ values. }
\label{fig:SFHNewReg}
\end{center}
\end{figure*}

\vspace{2mm} 

\noindent $\bullet$ First, it is noticeable from Figure \ref{fig:commonBins} that galaxies with rapidly increasing SFHs (purple/blue bins) and those with constant SFH (green bins) are clearly separated by a line that runs somewhat parallel to the SFS. As such, we first define (by eye) a line based on the upper boundary between the green and blue/purple bins at low stellar masses. This takes the form of log$_{10}$(sSFR) = -0.41$\times$ log$_{10}$(M$_{\star}$) -4.85, and is shown as the upper dark blue diagonal lines in Figures \ref{fig:slope} \& \ref{fig:SFHNewReg}. This line does not change with epoch. Sources above this line fall into our new region 1, and are deemed `SF increasing'. We see that while the distribution of points changes in this region as a function of time, galaxies with rapidly increasing SFHs always sit above this line. We also highlight again that there is a paucity of these sources at high stellar masses. \\

\noindent $\bullet$  Next, we consider points along the traditional SFS region. Here we find that at lower stellar masses this region contains almost exclusively systems that have constant SFHs (green bins), while at high stellar masses it contains galaxies with declining SFHs (black and red bins). The separation between these two types of galaxy qualitatively appears to trace M$^{*}_{\sigma-min}$ at all epochs (gold vertical lines). As such, we separate the traditional SFS region into two stellar mass ranges based on M$^{*}_{\sigma-min}$ (which hence evolves with redshift). To encapsulate the systems with constant SFH we define a line that traces the lower envelope of the sequence formed of green points as: log$_{10}$(sSFR) = -0.25$\times$log$_{10}$(M$_{\star}$) -7.10 at  M$^{*}<$M$^{*}_{\sigma-min}$. Then to bound the region which contains the bulk of the rapidly declining systems,  we define a line that extends from the intersection of the previous line and M$^{*}_{\sigma-min}$, to the log$_{10}$(sSFR)=-11.3\,yr$^{-1}$ and log$_{10}$(M$_{\star}$)=11.5\,M$_{\odot}$ point. This appears to bound the black/red bins at all epochs, but evolves in slope and normalisation with M$^{*}_{\sigma-min}$. These are shown as the central dark blue line in Figures \ref{fig:slope} \& \ref{fig:SFHNewReg} that breaks at M$^{*}_{\sigma-min}$. We define the lower stellar mass end of the SFS as region 2, called `SFS', and the higher stellar mass end as region 3, `Rapid quenching'. \\

\noindent $\bullet$  Next we find that as we move to lower sSFRs at all stellar masses, we enter a region dominated by galaxies with slowly declining SFHs (orange bins). Lower still the sample transitions to passive constant SFH sources (green bins) There is a clear separation between these populations by a line that takes the form of: log$_{10}$(sSFR) = -0.93$\times$log$_{10}$(M$_{\star}$) -1.3 and extends until it crosses the lower boundary of the SFS line at low stellar masses (log$_{10}$(M$_{\star}$)$\sim$8.5\,M$_{\odot}$). This line is shown by the lower diagonal dark blue line in Figure \ref{fig:slope} \& \ref{fig:SFHNewReg}. Above this line and below the SFS population we define our region 4, called `slow quenching'.  \\

\noindent $\bullet$  Finally, the low sSFR but constant SFH populations also appear to be relatively well separated by a dividing line at M$^{*}_{\sigma-min}$ (in Figure \ref{fig:slope}), and hence we split this population at this value. Defining two `passive' galaxy regions, one at low stellar masses (region 5, called 'Low mass passive?') and region 6 called `passive'. It is not apparently clear what the population region 5 is (hence the `?'). At low redshift this population actually defines the SFS, but by tracing the evolution of the populations through these panels, it can be seen as somewhat distinct from the SFS population at earlier times. For example, the bulk of the constant SFH population seen in the $0.1<z<0.25$ panel can be seen in the $0.4<z<0.55$ panel, but this is not the SFS in the earlier epoch. This population will be discussed extensively in paper II. \\        

In total this sub-division of the sSFR-M$_{\star}$ plane results in the 6 new regions which are shown in the bottom right panel of Figure \ref{fig:commonBins}. In summary, we define these regions to contain galaxies based on common recent star-formation history, labeled 1-6 as: star-formation increasing, the SFS, rapid quenching, slow quenching, low-mass passive, and passive.

\begin{figure*}
\begin{center}
\includegraphics[scale=0.35]{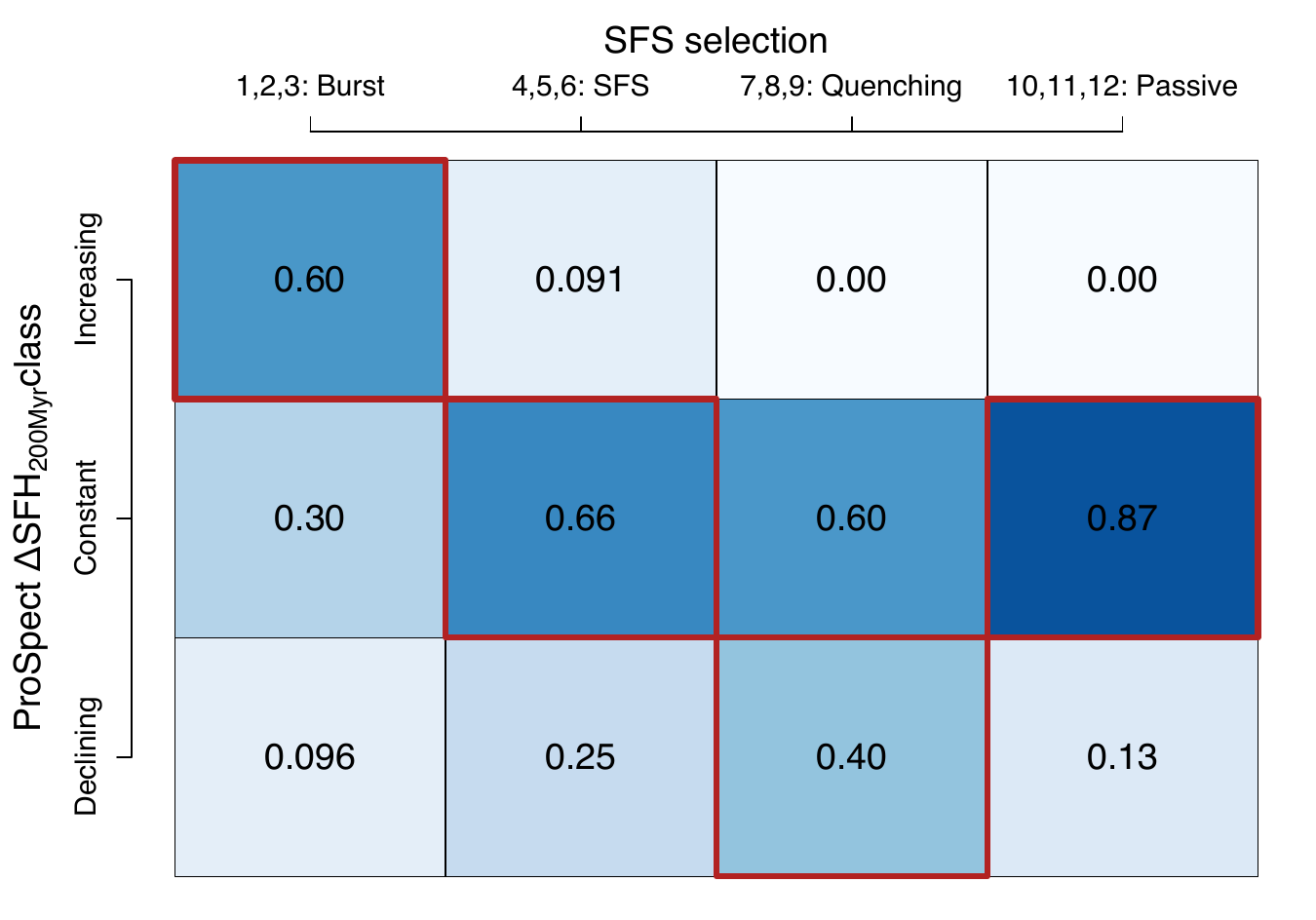}
\includegraphics[scale=0.35]{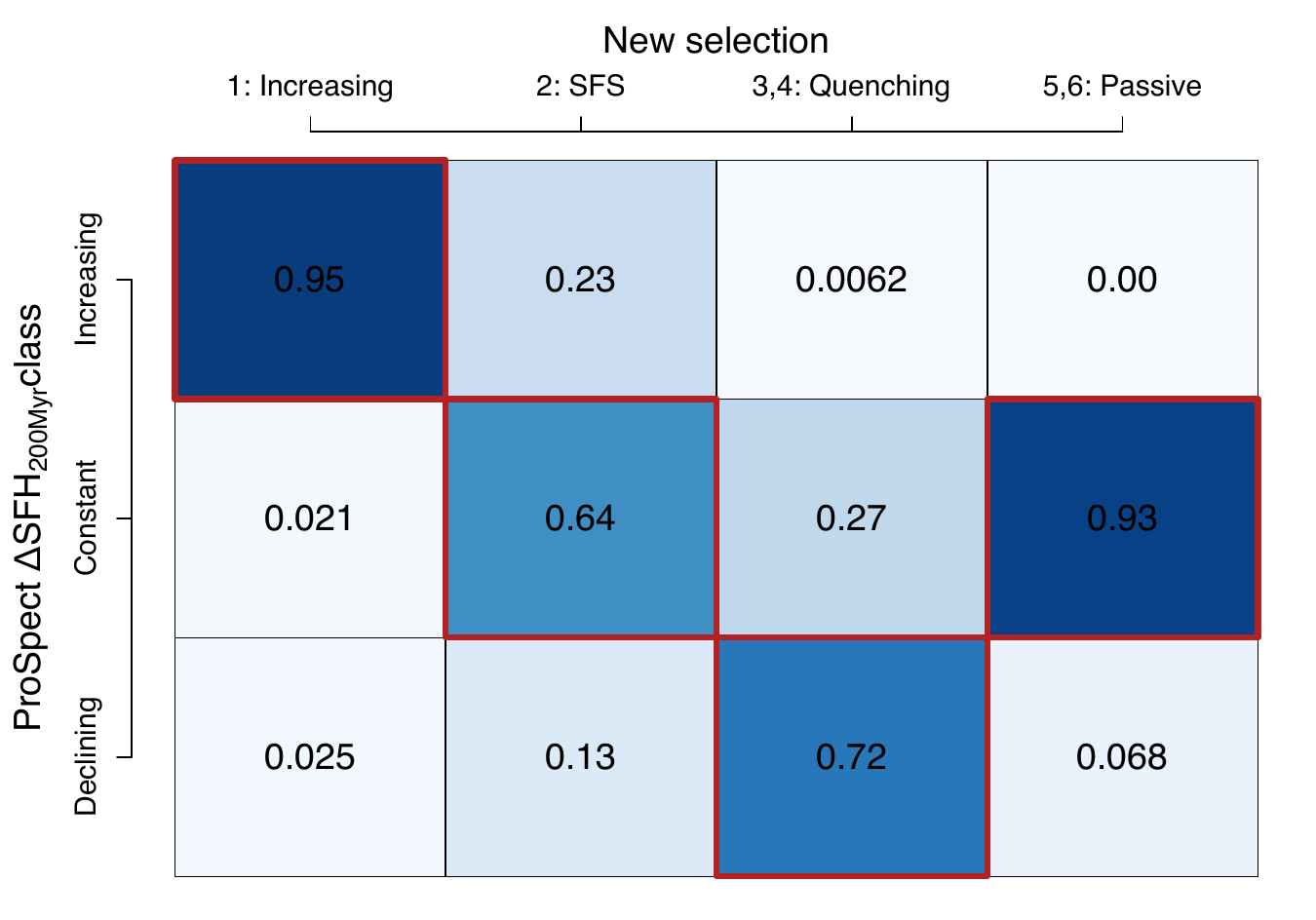}

\caption{The purity of our initial (based off Figure \ref{fig:selection}) selections using the position of objects relative to the SFS, left, compared to the purity of our new (based off Figure \ref{fig:SFHNewReg}) selections using the position of objects relative to the SFS and M$^{*}_{\sigma-min}$, right. Purity is measured as the fraction of objects in each selection region that have the same corresponding class as measured from their $\Delta$SFH$_{200\mathrm{\,Myr}}$ value. Bins that should have a high purity value if the selction robustly identifies galaxies with common $\Delta$SFH$_{200\mathrm{\,Myr}}$ values are highlighted in red. While all others should have a low purity value. We find that our new selection regions have higher purity values when selecting objects based off the \textsc{ProSpect}  $\Delta$SFH$_{200\mathrm{\,Myr}}$ value.  }
\label{fig:purity}
\end{center}
\end{figure*}

To determine if these regions robustly select for galaxies with common SFHs, the left column of Figure \ref{fig:SFHNewReg} shows a compilation of the \textsc{ProSpect} SFHs for galaxies in each selection region, with all epochs combined into a single figure. The coloured lines show the median SFH and interquartile range. Qualitatively we can see that the selection regions are identifying galaxies with common SFHs. This is also reflected in the distribution of $\Delta$SFH$_{200\mathrm{\,Myr}}$ values (shown in the right column), which are as expected given how the selection regions are defined. This suggests that these regions do in fact select for galaxies that have common recent SFHs. It is therefore interesting that a number of these regions do not significantly change with redshift. However, we leave the bulk of the discussion around the physical interpretation of these results to the second paper in this series.

To provide a more quantitative metric of the robustness of our new selection regions, we also compare the purity of these selections to those using the selections based off the position of galaxies in relation to the SFS, used at the start of this work.  To directly compare, we combine the quenching galaxies in our new selections (regions 3 and 4) into a single class, and our passive galaxies (regions 5 and 6) into a single class. We then also combine the selection regions in Figure \ref{fig:selection} into burst (regions 1,2,3), SFS (regions 4,5,6), quenching (regions 7,8,9) and passive (regions, 10,11,12). Figure \ref{fig:purity}, then shows the fraction of galaxies in each class that have the correct corresponding $\Delta$SFH$_{200\mathrm{\,Myr}}$ assignment as either increasing (burst), constant (SFS and passive), or declining (quenching). The left panel displays the old selections, based off Figure \ref{fig:selection}, while the right panel shows the new selections based off Figure \ref{fig:SFHNewReg}. We also highlight the boxes where you would like to see a high purity fraction, if the selections are robustly identifying regions with common $\Delta$SFH$_{200\mathrm{\,Myr}}$ values. All other boxes should have low purity fraction.   

We find that our new selection regions have a similar or higher purity value for selection regions with common $\Delta$SFH$_{200\mathrm{\,Myr}}$ values, and therefore likely better select for sources with common SFHs. Most notably, this is true for sources which would be selected as 'Quenching' via each method.  For the original SFS selection, galaxies in a traditional `Quenching region' (below the SFS), actually show a significant fraction of galaxies that have constant SFHs; this is significantly reduced for the new selections. While the exact values of the purity in these cases will depend on the fine details of the selections used, this figure highlights that our new selection methodology does in fact better select for regions with common $\Delta$SFH$_{200\mathrm{\,Myr}}$ values, than using position relative to the SFS alone.

\begin{figure*}
\begin{center}
\includegraphics[scale=0.55]{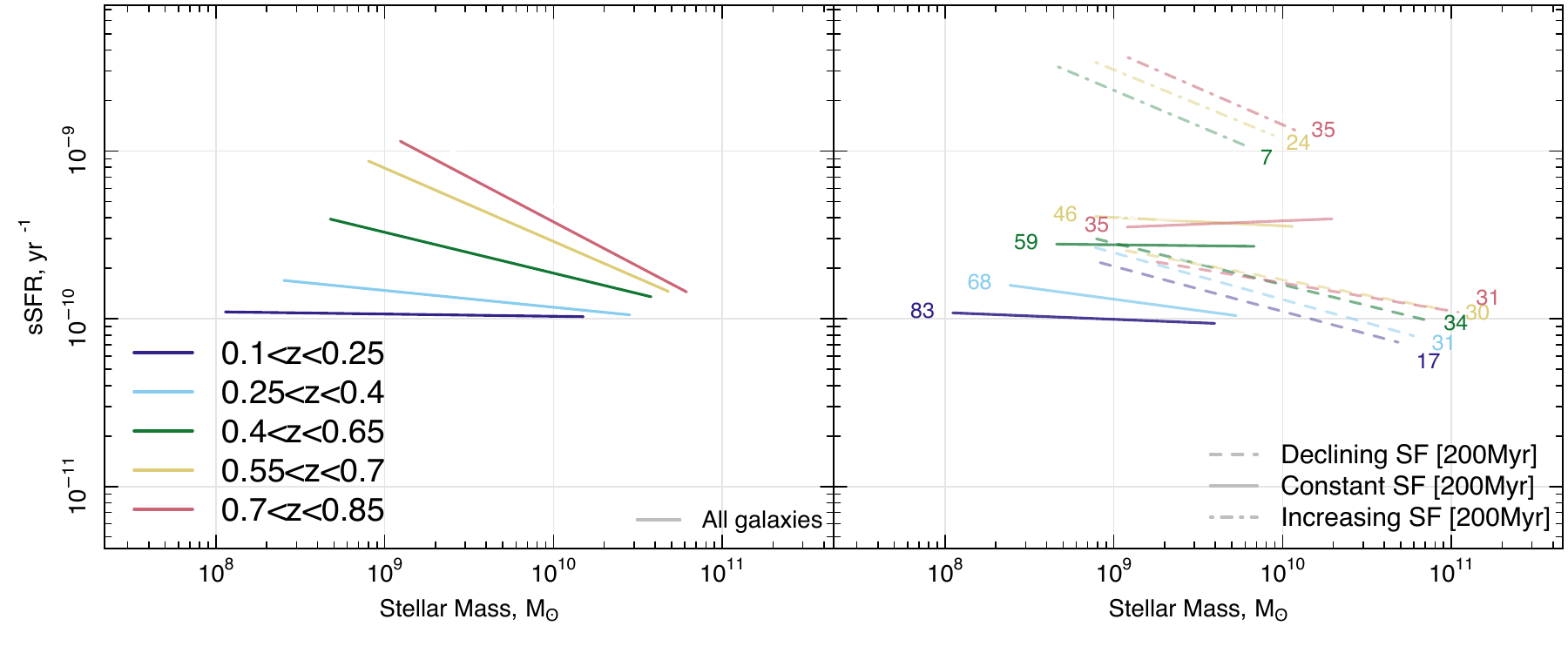}

\caption{Fits for the evolution of the slope and normalisation of the SFS using samples selected on their recent change in SFH over the last 200\,Myr. Left panel shows linear fits to the full sample at each epoch. In the right panel we split our sample into \textit{declining} ($\Delta SFH_{200Myr}<-0.1$), \textit{constant} ($-0.1<\Delta SFH_{200Myr}<0.1$) and \textit{increasing} ($\Delta SFH_{200Myr}>0.1$). Lines are alpha transparent based on the number of galaxies that fall in each sample at given epoch (there is no line for the increasing population at $0.1<z<0.4$  as there are only a very small number of increasing galaxies in our sample at the epoch - 0 and 1 at each epoch respectively).  Lines are plotted over the stellar mass range that contains 90\% of the respective sample (0.05-0.95 quantile). Numbers associated with each line are the percentage of galaxies that fall within that sample at the given epoch. We find that the slope of each population is different but stays roughly the same at each epoch - the \textit{increasing} population has a steep negative slope, the \textit{constant} population a flat slope, and the \textit{declining} population a shallow negative slope. The populations evolve differently in normalisation, with the \textit{constant} population evolving significantly since $z=0.85$, the \textit{increasing} population showing moderate evolution and the \textit{declining} population showing little normalisation evolution. We also see that the fraction of galaxies in the \textit{declining} population remains relatively constant, while the fraction in the \textit{increasing} population goes down and the \textit{constant} population goes up. In combination, differences in slope, normalisation evolution and fraction of galaxies in each population leads to the observed evolution in the slope and normalisation of the full SFS (left). At higher redshifts the steep-slope \textit{increasing} population dominates the galaxy distribution, leading to a steep SFS. As we move to later times, the distribution is dominated by the flat-slope \textit{constant} population, leading to a much flatter SFS. }
\label{fig:SlopesEvol}
\end{center}
\end{figure*}

\subsection{The evolution of galaxies through the sSFR-M$_{\star}$ plane, the SFS and consequences for the overall distribution of galaxies}
\label{sec:planeEvol}

Throughout this paper we have alluded to the fact that our results indicate the sSFR-M$_{\star}$ plane (and the SFS) are evolving in a complex manner. In this final section we aim to use the results derived in this work to explore how (and why) the plane is evolving from $z\sim1$ to the present day and to explain some of the key observational features of the SFS. 

The evolution of the sSFR-M$_{\star}$ plane is typically defined by the changing shape and normalisation of the star-forming sequence \citep[$e.g.$][]{Whitaker14, Schreiber15, Davies16b, Lee15, Leslie20, Thorne21}. While the details of these works differ, the overall consensus is that the SFS (in terms of sSFR) declines in normalisation and flattens over time. The other key feature of the SFS is a steepening of the relation at the high stellar mass end (flattening in SFR-M$_{\star}$, the so-called `turn-over mass'). The stellar mass at which the SFS changes slope has been traced robustly by \cite{Thorne21}, and other authors, and found to move to lower stellar masses as the Universe evolves. There have been many physical interpretations of this `turn-over' point in the SFS from the fact that the SFS only holds true for disks and high stellar mass galaxies have significant non-star-forming bulge components \citep{Cook20} to variations in star-formation efficiency at high stellar masses \citep{Lee15}, and efficient quenching in high mass halos \citep{Leauthaud12} attributed to AGN feedback \citep{Springel05}, tidal disruptions/mergers \citep{Hopkins06}, the impact of dense environments on gas in galaxies \citep{Peng10,Brown17} and/or variations to the mode of gas accretion \citep{Keres05, Nelson13}. However, as we point out in D22 the `turn-over' in the SFS measured by \cite{Thorne21} also closely traces the evolution of M$^{*}_{\sigma-min}$. If this is the case, could they both have a common origin? This is explored in terms of just the distributions alone (not physical origin) in Figure 10 of D22, where we show that the SFS has very different slopes above and below M$^{*}_{\sigma-min}$. We can now also see this quantitatively in Figure \ref{fig:slope} where the populations along the traditional SFS region transitions from galaxies with constant SFH to those with rapidly declining SFHs at M$^{*}_{\sigma-min}$. At this point we also see a change in slope of the SFS. This immediately suggests the turn-over of the SFS represents the transition between galaxies that are evolving in a constant self-regulated manner at low stellar masses and those that are rapidly quenching at higher stellar masses.

To explore this further, we first consider whether or not the changing shape and normalisation of the SFS is a consequence of galaxies in different regions of sSFR-M$_{\star}$ plane having different SFHs, and the relative contribution of galaxies in each region changing as a function of time. For example, does the SFS for galaxies with similar SFHs evolve in a self-similar way, but the overall SFS change due to the dominance of specific SFH types? In the left panel of Figure \ref{fig:SlopesEvol} we first show the evolution of a linear fit to the SFS population taken from Figure \ref{fig:selection}. Note that this only fits below the turnover mass \textcolor{black}{for star-forming galaxies determined in \cite{Thorne21}}.  Here we see the characteristic change in slope and normalisation as the Universe evolves, where the slope becomes flatter and declines in normalisation. In the right panel we now split the population based on $\mathrm{\Delta SFH_{200\,Myr}}$ value into \textit{declining} ($\mathrm{\Delta SFH_{200Myr}<-0.1}$), \textit{constant} ($\mathrm{-0.1<\Delta SFH_{200Myr}<0.1}$) and \textit{increasing} ($\mathrm{\Delta SFH_{200Myr}>0.1}$) and fit each population separately with a linear relation. We plot these lines over the stellar mass range that contains 90\% of the respective sample (0.05-0.95 quantile) to highlight where these sources lie, and also include labels for the fraction in galaxies in each selection at each epoch. What is clear from this figure, is that galaxies with similar SFHs evolve in very similar ways in the sSFR-M$_{\star}$ plane. The galaxies that are increasing in star formation have a steep slope that evolves moderately in normalisation but their numbers decrease rapidly with time (the fraction in the two lowest redshift bins is $\sim0$, hence they are not shown), the constant SFR population has a flat slope that evolves strongly in normalisation as the numbers in this population grow rapidly, and the galaxies that are decreasing in star formation have a moderate slope and evolve very little in normalisation, but decrease in number. 

We can now also see how the varying contributions of these populations shape the overall evolution of the SFS. At the high redshift end the full sample has a large contribution from both the increasing and declining SFH populations. This produces a steep slope with high normalisation. As the universe evolves galaxy populations shift to being dominated by those with constant SFHs, causing the overall slope to flatten. We also see the point at which the constant SFH population overlaps with the declining population move to lower stellar masses as the universe evolves. This traces the turnover point in the SFS. As such, it appears that it is the varying contribution and different normalisation evolution of these SFH classes that ultimately leads to changes in the shape and normalisation evolution of the SFS, with the turnover mass being the boundary between sources with constant SFH and those with declining SFHs.

\vspace{2mm}

\begin{figure*}
\begin{center}
\includegraphics[scale=0.13]{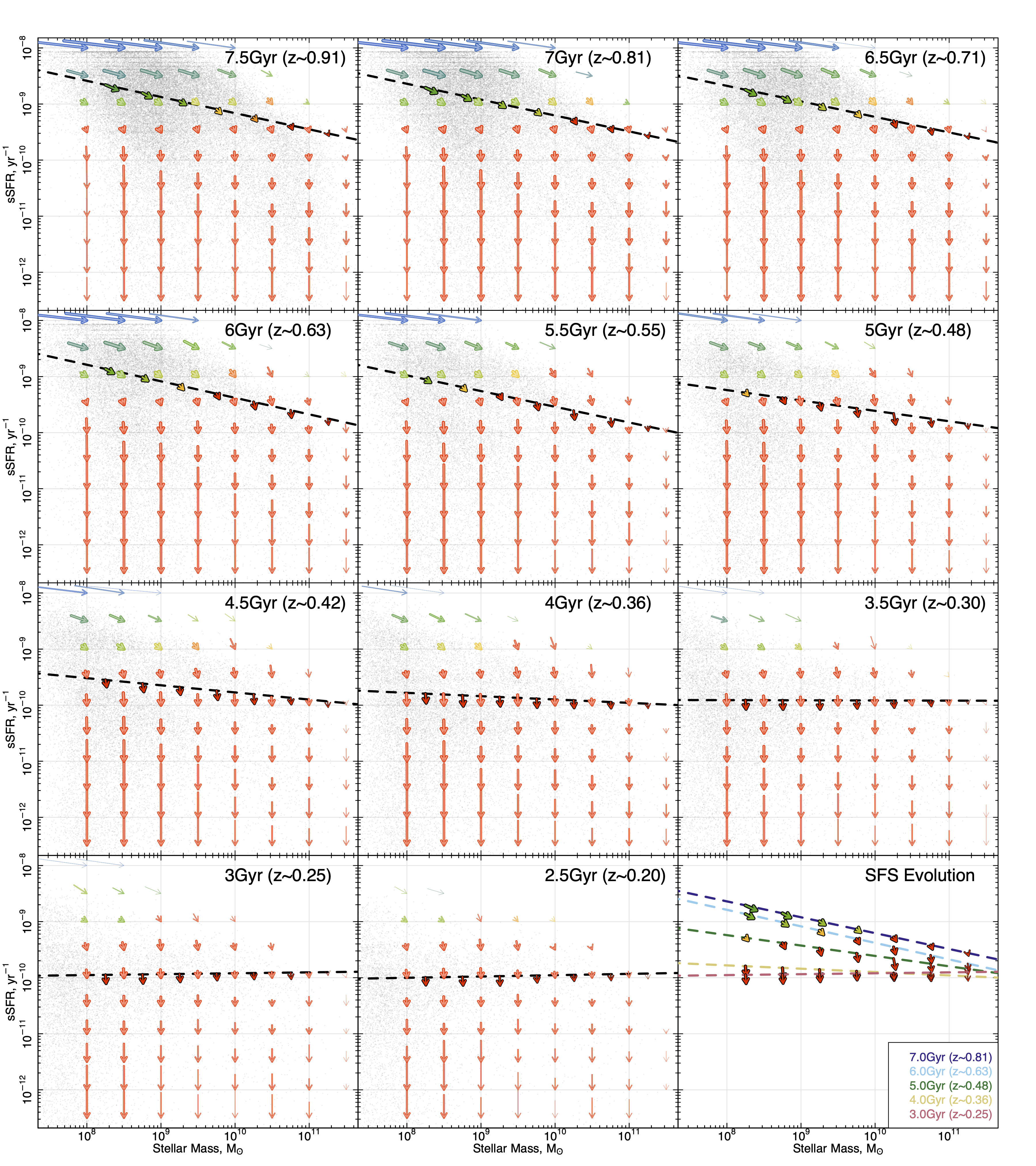}

\caption{The evolving distribution of galaxies in the sSFR-M$_{\star}$ plane. In each plane we show the t=0 distribution of galaxies at a given lookback time $\pm$0.4\,Gyr as the grey points. The dashed line shows a fit to the SFS at this epoch. The sSFR-M$_{\star}$ plane is then binned in log$_{10}$(M$_{\star}$) and log$_{10}$(sSFR). In each bin the median evolution of galaxies over the past 200\,Myr is plotted as an arrow. The colour of the arrows also show the slope of the evolution, and the weight of the arrow is representative of log$_{10}$(\#) of galaxies in the bin. Thus, galaxies with a colour the same as the SFS fits are evolving perpendicular to the SFS. Black-bordered arrows show the same but for galaxies which sit on the SFS at a given epoch. These black arrows define how the SFS is evolving at a given epoch and stellar mass, with the bottom right panel only showing the evolution of SFS galaxies at all epochs.}
\label{fig:PlaneEvol}
\end{center}
\end{figure*}

Taking this analysis even further, it is possible to explore how the full sSFR-M$_{\star}$ plane is evolving with time, not just the SFS. To do this, we split our full sample into 11 look-back time bins from 0 to 8\,Gyr with $\Delta t=0.5$\,Gyr. We opt to use look-back time here over redshift as we wish to explore the time evolution of the plane. In Figure \ref{fig:PlaneEvol}, the grey data points show the sSFR-M$_{\star}$ distribution at each time step. We then fit the relation linearly to give the SFS at each epoch (dashed black lines).  

We then take points along the SFS in log$_{10}$(M$_{\star}$/M$_{\odot}$)=0.5 bins and select all galaxies within a $\Delta$log$_{10}$(M$_{\star}$/M$_{\odot}$)=0.5 and $\Delta$log$_{10}$(sSFR/Gyr)=0.5 region around the points. We then use each galaxy's SFH to calculate the median sSFR and stellar mass change over the last 200\,Myr. This gives the median trajectory of galaxies through the sSFR-M$_{\star}$ plane at that point. We then repeat this process for an equally spaced grid of $\Delta$log$_{10}$(M$_{\star}$/M$_{\odot}$)=0.5 and $\Delta$log$_{10}$(sSFR/Gyr)=0.5 bins across the full sSFR-M$_{\star}$ plane. 

In Figure \ref{fig:PlaneEvol} we show these trajectories as arrows, where the colour of the arrow represents the gradient (blue=predominantly growing in stellar mass, red=predominantly declining in sSFR) and the weight of the arrow line is proportional to log(number of sources) in each bin. The points along the SFS at each epoch are bounded by black lines and a compendium of just some of the SFS lines and arrows is given in the bottom right panel to display the overall evolution of the SFS. While we show the sSFR-M$_{\star}$ plane here, we also produce a similar figure for the SFR-M$_{\star}$ plane, which is given in the appendix. 

These Figures display the overall evolution of the plane as it shifts from intense star-formation and growing rapidly in stellar mass at early times, to largely declining in sSFR and growing little in stellar mass at the current epoch. Interestingly, for our \textsc{ProSpect} SFHs at least, the trajectory of galaxies through this plane is largely dependant on the position within the plane, which does not change with epoch. However, the relative numbers of galaxies at each region of the plane does change with time shifting the overall distribution. $i.e.$ on average a galaxy's recent change in SFH is dependent on its stellar mass and sSFR alone, but the numbers of galaxies at each stellar mass and sSFR change over time. 

It is also interesting that we see a somewhat smooth evolution of galaxy trajectories that matches the evolution of the SFS and the overall population. We remind the reader that the position of the grey points in this diagram and therefore the fitted SFS are not derived using the galaxy's SFH (they are in-situ measured values), such that a smooth transition of the SFS from one panel to the next that follows the trajectory of the arrows is not a given. One could attempt here to back propagate the local population to earlier epoch to explore if this back-propagated population matches the observed distribution at an earlier time \citep[$e.g.$ see ][]{Sanchez18, Iyer18, Phillipps19,Bellstedt20}. However, this is the subject of a future paper (Davies et al in prep) and as such we do not discuss it here. 

This figure also displays why the SFS evolves in the observed manner. We see a rapid growth in stellar mass of the low stellar mass and high sSFR points at high redshift, which moves these galaxies to the high stellar mass end and they then begin to decline in star-formation \textcolor{black}{(these are the systems with rapidly increasing star formation in $e.g.$ Figure \ref{fig:z3_rel} panel 1)}. This low stellar mass and high sSFR population is not replenished and the low stellar mass end becomes increasing dominated by low sSFR points, leading to a rapid change in the slope and normalisation at the low stellar mass end. Conversely, at the high stellar mass end galaxies are not growing in stellar mass and declining slowly in sSFR. There we see much slower evolution in the normalisation of the SFS. As time evolves, both the low and high stellar mass end reach a point where all galaxies are slowly declining in SFR, the SFS is flat and will continue to decline into the future. While not in the sample explored here, the strongly star-forming population will now be predominantly found at even lower stellar masses (downsizing). Here we are witnessing in-situ and across the SFR-M$_{\star}$ plane at M$_{\star}>10^{8}$M$_{\odot}$, the transition of galaxies from a population that is dominated by actively star-forming systems, to a passively evolving population that will slowly decline in star-formation as the Universe ages - as has previously been seen in the cosmic star-formation history (CSFH). The Universe truely is on the decline and as we move into the future this trend will continue with lower and lower stellar mass systems joining the passively declining population.

\section{Summary and Conclusions}

In this work we have explored the variation in galaxy SFHs across the sSFR-M$_{\star}$ plane to determine how the overall shape and distribution of galaxies is evolving with time. We first divide the plane into subregions based on stellar mass and sSFR, and highlight that there is strong variation in SFH between subregions. We then show that the SFHs and the number of galaxies in each sub-region evolves strongly with time, leading to a complex evolution of the sSFR-M$_{\star}$ plane. We highlight that the traditional method of selecting galaxies with common SFHs using position relative to the SFS my be flawed, as the distribution of SFHs at similar positions relative to the SFS changes with stellar mass. 

We define a new single metric for parametrising the recent SFH of galaxies,  $\Delta$SFH$_{200\mathrm{\,Myr}}$,  and use this to show that galaxies with common SFHs populate similar regions of the sSFR-M$_{\star}$ plane. We use these to define new sub-regions of the sSFR-M$_{\star}$ plane and show that they do in fact select galaxies with common $\Delta$SFH$_{200\mathrm{\,Myr}}$ values. Finally, we go on to explore the overall evolution of the sSFR-M$_{\star}$ plane, and show that it is the varying contribution of galaxies with different recent SFHs that leads to the evolution of slope, normalisation and turn-over mass. We suggest that the evolution of the turn-over mass and the minimum sSFR dispersion, M$^{*}_{\sigma-min}$, point is likely due to the boundary point between galaxies with constant, self-regulated star formation, and those with rapidly declining star-formation. The physical processes that may drive this transition will be explored in the next paper in this series. 

We note that this first paper largely defines the observational trends that we observe with regards to the variation of SFHs across the sSFR-M$_{\star}$ plane, their evolution and how this leads to the observed trends in the evolution of the SFS and sSFR-M$_{\star}$ plane. We leave the detailed analysis of the mechanisms driving these trends and their astrophysical interpretation to paper II.

\section*{Acknowledgements}

LJMD, ASGR, and SB acknowledge support from the Australian Research Councils Future Fellowship scheme (FT200100055 and
FT200100375). JET was supported
by the Australian Government Research Training Program (RTP)
Scholarship. MS acknowledge support from the Polish National Agency for Academic Exchange (Bekker grant BPN/BEK/2021/1/00298/DEC/1), the European Union's Horizon 2020 Research and Innovation programme under the Maria Sklodowska-Curie grant agreement (No. 754510). Parts of this research were
conducted by the Australian Research Council Centre of Excellence
for All Sky Astrophysics in 3 Dimensions (ASTRO 3D), through
project number CE170100013. DEVILS is an Australian project
based around a spectroscopic campaign using the Anglo-Australian
Telescope. DEVILS is part funded via Discovery Programs by the
Australian Research Council and the participating institutions. The
DEVILS website is \url{devils.research.org.au}. The DEVILS data are
hosted and provided by AAO Data Central (\url{datacentral.org.au}).

\section{Data Availability}

Data products used in this paper are taken from the internal DEVILS
team data release and presented in \cite{Davies21} and \cite{Thorne21}. These catalogues will be made public as part DEVILS
first data release described in Davies et al. (in preparation).

\appendix

\section{Definition of the regions defined in Section 3.1}

Included here are two tables detailing the selections described in Section \ref{sec:var} and Figure \ref{fig:selection}, such that they can be reproduced if required. These selection are used to identify different regions of the SFR-M$_{\star}$ plane based on stellar mass relative to M$^{*}_{\sigma-min}(z)$ and the SFS.  

\begin{table*}
\caption{Reference table between numbered regions in Figure \ref{fig:selection} and main text, and the selections they contain.}
\begin{center}
\begin{tabular}{c c c c c}
Region & Stellar Mass & sSFR  & Stellar Mass &  sSFR \\
Number & Region & Region & Selection & Selection \\ 
\hline
1 & Low Mass & Bursting & M$_{lim}$ $<$ M$_{\star}$ $<$ M$^{*}_{\sigma-min}(z)$-0.3\,dex &  sSFR$>$SFS+0.3\,dex\\
2 &  M$^{*}_{\sigma-min}$  & Bursting & M$^{*}_{\sigma-min}(z)$-0.3\,dex $<$ M$_{\star}$ $<$ M$^{*}_{\sigma-min}(z)$+0.3\,dex &  sSFR$>$SFS+0.3\,dex\\
3 & High Mass  & Bursting & M$^{*}_{\sigma-min}(z)$+0.3\,dex  $<$ M$_{\star}$ $<10^{11.5}$\,M${\odot}$  &  sSFR$>$SFS+0.3\,dex\\
4 & Low Mass  & Sequence & M$_{lim}$ $<$ M$_{\star}$ $<$ M$^{*}_{\sigma-min}(z)$-0.3\,dex &  SFS-0.3\,dex<sSFR$<$SFS+0.3\,dex\\
5 & M$^{*}_{\sigma-min}$& Sequence  & M$^{*}_{\sigma-min}(z)$-0.3\,dex $<$ M$_{\star}$ $<$ M$^{*}_{\sigma-min}(z)$+0.3\,dex  &  SFS-0.3\,dex<sSFR$<$SFS+0.3\,dex\\
6 & High Mass & Sequence &  M$^{*}_{\sigma-min}(z)$+0.3\,dex  $<$ M$_{\star}$ $<10^{11.5}$\,M${\odot}$  &  SFS-0.3\,dex<sSFR$<$SFS+0.3\,dex\\
7 & Low Mass & Quenching & M$_{lim}$ $<$ M$_{\star}$ $<$ M$^{*}_{\sigma-min}(z)$-0.3\,dex &  SFS-0.9\,dex<sSFR$<$SFS-0.3\,dex\,dex\\
8 & M$^{*}_{\sigma-min}$ & Quenching & M$^{*}_{\sigma-min}(z)$-0.3\,dex $<$ M$_{\star}$ $<$ M$^{*}_{\sigma-min}(z)$+0.3\,dex  &  SFS-0.9\,dex<sSFR$<$SFS-0.3\,dex\\
9 & High Mass & Quenching & M$^{*}_{\sigma-min}(z)$ +0.3\,dex $<$ M$_{\star}$ $<10^{11.5}$\,M${\odot}$  &  SFS-0.9\,dex<sSFR$<$SFS-0.3\,dex\\
10 & Low Mass & Passive & M$_{lim}$ $<$ M$_{\star}$ $<$ M$^{*}_{\sigma-min}(z)$-0.3\,dex  &  sSFR$<$SFS-0.9\,dex \\
11 & M$^{*}_{\sigma-min}$ & Passive & M$^{*}_{\sigma-min}(z)$-0.3\,dex $<$ M$_{\star}$ $<$ M$^{*}_{\sigma-min}(z)$+0.3\,dex  &  sSFR$<$SFS-0.9\,dex \\
12 & High Mass & Passive & M$^{*}_{\sigma-min}(z)$+0.3\,dex $<$ M$_{\star}$ $<10^{11.5}$\,M${\odot}$  &  sSFR$<$SFS-0.9\,dex \\

\end{tabular}
\end{center}
\label{tab:def}
\end{table*}

\begin{table*}
\caption{Characteristics of galaxies in each selection region, relocated here for reference. Region numbers correspond to those in Table \ref{tab:def} and Figure \ref{fig:selection}. Redshifts are the same ranges used elsewhere in this work from low to high, $e.g. z1: 0.1<z<0.25, z5: 0.7<z<0.85$. }
\begin{center}
\begin{tabular}{c | c c c c c | c c c c c | c c c c c} 
& \multicolumn{5}{c}{Median log$_{10}$(Stellar Mass)} & \multicolumn{5}{c}{Median log$_{10}$(sSFR)}   & \multicolumn{5}{c}{\# of galaxies in region}   \\
Region  & $z1$ & $z2$ & $z3$ & $z4$ & $z5$ & $z1$ & $z2$ & $z3$ & $z4$ & $z5$ & $z1$ & $z2$ & $z3$ & $z4$ & $z5$ \\
\hline
1 & 8.34 & 8.68 & 8.98 & 9.21 & 9.41 & -9.38 & -9.08 & -8.89 & -8.62 & -8.52 & 247 & 815 & 1425 & 2241 & 2796 \\ 
2 & 9.18 & 9.42 & 9.67 & 9.93 & 10.24 & -9.46 & -9.31 & -9.09 & -8.96 & -9.00 & 66 & 311 & 315 & 345 & 215 \\ 
3 & 9.78 & 10.02 & 10.33 & 10.55 & 10.80 & -9.46 & -9.39 & -9.29 & -9.29 & -9.31 & 65 & 135 & 69 & 71 & 49 \\ 
4 & 8.41 & 8.74 & 9.00 & 9.20 & 9.43 & -9.91 & -9.74 & -9.35 & -9.10 & -9.03 & 1206 & 1629 & 1677 & 2759 & 2954 \\ 
5 & 9.17 & 9.49 & 9.77 & 10.05 & 10.32 & -9.92 & -9.75 & -9.51 & -9.43 & -9.44 & 515 & 875 & 837 & 1017 & 943 \\ 
6 & 9.93 & 10.19 & 10.41 & 10.66 & 10.87 & -9.95 & -9.89 & -9.78 & -9.78 & -9.76 & 368 & 837 & 560 & 611 & 448  \\ 
7 & 8.36 & 8.72 & 8.94 & 9.20 & 9.41 & -10.39 & -10.25 & -10.01 & -9.72 & -9.63 & 1055 & 2128 & 1397 & 1344 & 1605 \\ 
8 & 9.14 & 9.44 & 9.76 & 10.10 & 10.38 & -10.37 & -10.31 & -10.14 & -10.01 & -10.06 & 304 & 775 & 408 & 511 & 574 \\ 
9 & 10.09 & 10.31 & 10.52 & 10.73 & 10.90 & -10.41 & -10.39 & -10.31 & -10.30 & -10.31 & 229 & 674 & 511 & 495 & 395 \\ 
10 & 8.26 & 8.65 & 8.92 & 9.17 & 9.42 & -13.96 & -12.74 & -11.86 & -11.39 & -10.99 & 2266 & 4048 & 3849 & 3372 & 3259 \\ 
11 & 9.12 & 9.43 & 9.72 & 10.06 & 10.39 & -12.06 & -11.79 & -11.54 & -11.36 & -11.34 & 275 & 966 & 715 & 867 & 1243 \\ 
12 & 10.36 & 10.47 & 10.59 & 10.74 & 10.91 & -12.18 & -11.89 & -11.77 & -11.68 & -11.61 & 479 & 1448 & 1135 & 1153 & 1073 \\ 

\end{tabular}
\end{center}
\label{tab:subsamp}
\end{table*}

\section{Distribution of SFH variation across the plane using different SFH scalings}
\label{app:log10} 

\textcolor{black}{In this paper we opt to use a change in linear SFR as our metric for the recent change in SFH in order to capture the direct change in SFR in a galaxy (not relative to current SFR or any other quantity). However, as noted in the text, one could equally use change in log$_{10}$(SFR), or sSFR, or log$_{10}$(sSFR), as a similar diagnostic metric. Each has its benefits and drawbacks, and is somewhat defined by the specific question being addressed.  For transparency, in this appendix we also briefly explore using change in log$_{10}$(SFR), or sSFR, or log$_{10}$(sSFR) as our metric for recent change in SFH, and discuss differences in some of the key figures in this paper) and how that might affect any inferences drawn in this work.} 

\begin{figure*}
\begin{center}
\includegraphics[scale=0.105]{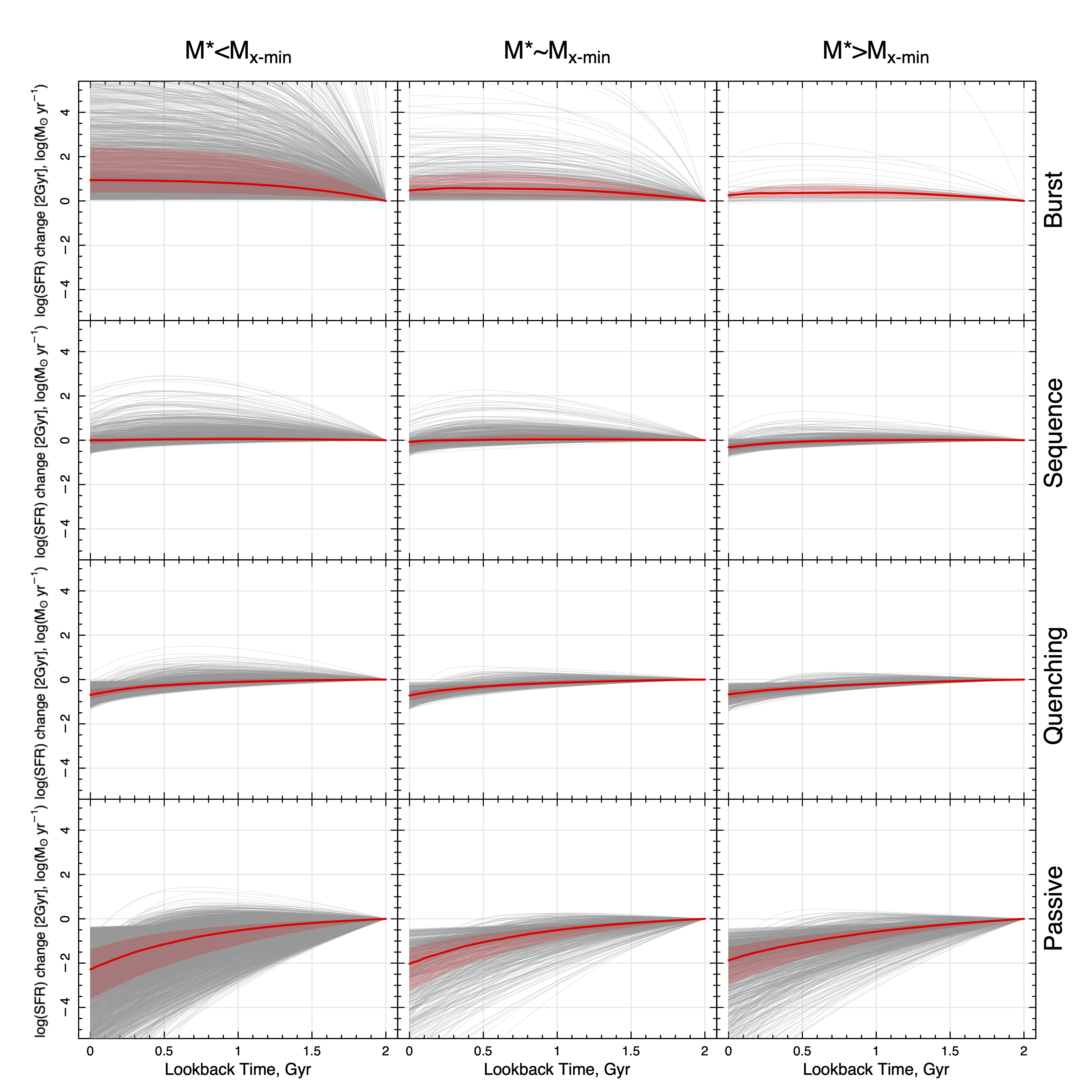}
\caption{The same as Figure \ref{fig:z3_rel}, but now showing SFHs as log$_{10}$(SFR). We still see strong variation of SFHs across the different regions, selected with respect to position within the  sSFR-M$_{\star}$ plane.  }
\label{fig:log_SFHs_z3Rel}
\end{center}
\end{figure*}

\textcolor{black}{First we reproduce Figure \ref{fig:z3_rel} but showing the SFHs as log$_{10}$(SFR) in Figure \ref{fig:log_SFHs_z3Rel}. Note we have restricted the y-axis range here such that we can see smaller changes in the general trend, at the expense of outlying SFHs. We still see strong variation in SFHs across the sSFR-M$_{\star}$ plane, with galaxies sitting above the SFS having significant rises in their star-formation over the last 2\,Gyr, SFS galaxies showing relatively flat SFHs, and galaxies in the `quenching' region showing declining SFRs. The main difference we see with respect to Figure \ref{fig:z3_rel} is that the most rapidly declining galaxies in log$_{10}$(SFR) are those with the lowest sSFRs in the `passive' regions. This is expected, as in these regions just a small change in linear SFR, can result in a large change in log$_{10}$(SFR). We do also see some (albeit weaker) trends with stellar mass, such as the high stellar mass galaxies within the SFS population having slightly declining SFRs compared to their lower stellar mass counterparts - similar to the linear SFHs in Figure \ref{fig:z3_rel}. Finally, we note that there does appear to be more variation/stochasticity in SFHs in lower stellar mass galaxies, particularly in the `burst' region, which may be indicative of star-bursting populations leading to a increased diversity in recent SFHs - similar to that see in the linear SFHs.}

\begin{figure*}
\begin{center}
\includegraphics[scale=0.2]{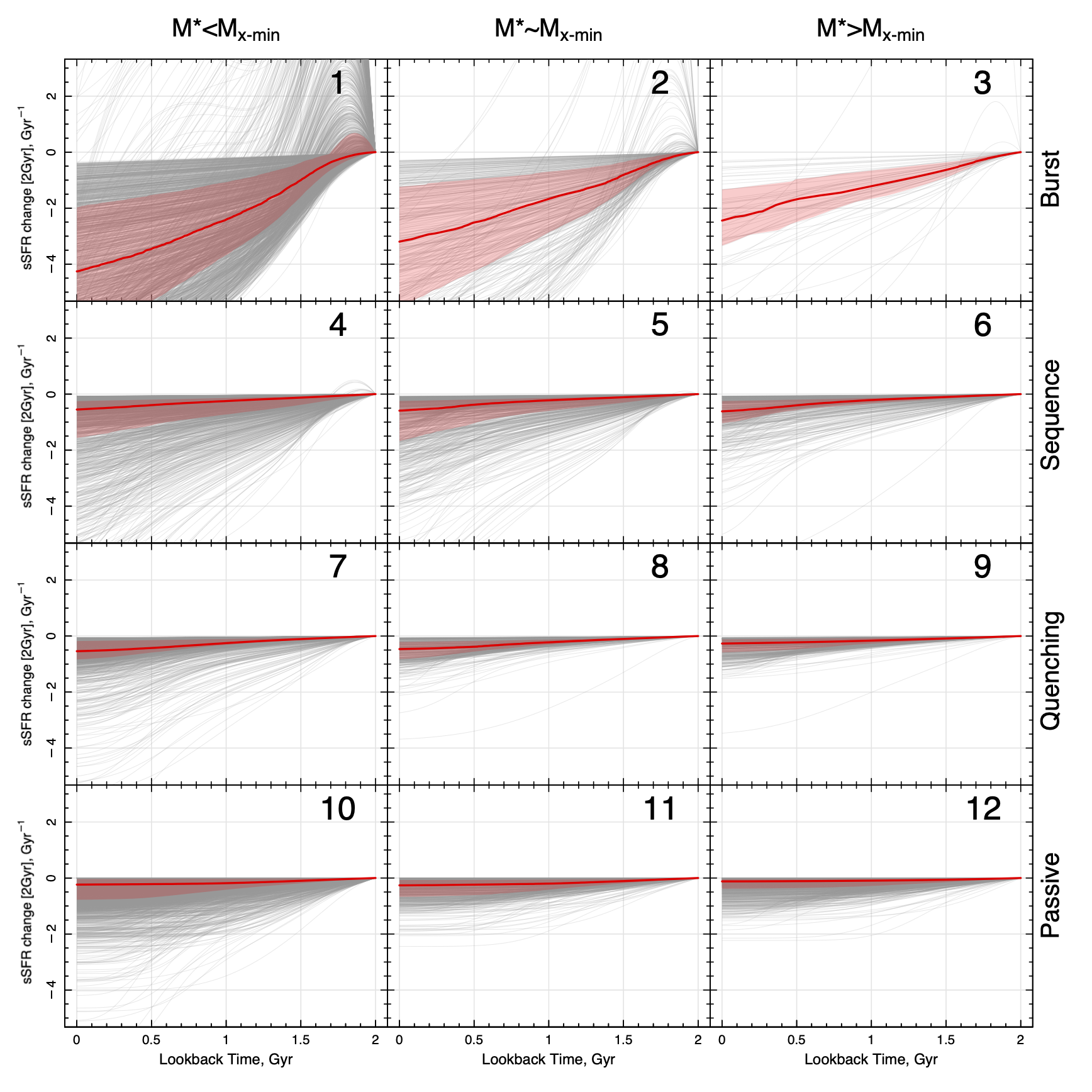}
\caption{The same as Figure \ref{fig:z3_rel}, but now showing SFHs as linear sSFR - note that units are now in Gyr$^{-1}$ for ease of plotting. We still see strong variation of SFHs across the different regions, selected with respect to position within the  sSFR-M$_{\star}$ plane.  }
\label{fig:linear_specific_SFHs_z3Rel}
\end{center}
\end{figure*}

\textcolor{black}{Next, we again reproduce Figure \ref{fig:z3_rel} but now showing the SFHs as linear sSFR in Figure \ref{fig:linear_specific_SFHs_z3Rel}. In this process, we calculate the sSFR of the galaxy at each time step along its SFH (taking into account mass recycling) and display the change in sSFR over the last 2\,Gyr. While we once again sees strong variation in SFHs across the sSFR-M$_{\star}$ plane, the most striking difference in comparison to Figure \ref{fig:z3_rel} is that all galaxies show declining sSFRs. This is due to the fact that galaxies with anything but very extremely increasing SFRs will grow more rapidly in stellar mass than their SFR is increasing, leading to declining sSFRs ($i.e.$ for a constant SFR, the galaxy grows in stellar mass and sSFR goes down). Interestingly, but not surprisingly, it is the galaxies with the highest sSFRs but lowest stellar mass that show the the most rapidly declining sSFHs as they are gaining stellar mass rapidly in comparison to their exisiting stellar mass. Again, we do also see potentially larger diversity in SFHs at the low stellar mass end, likely indicative of stochasticity.   }           

\begin{figure*}
\begin{center}
\includegraphics[scale=0.195]{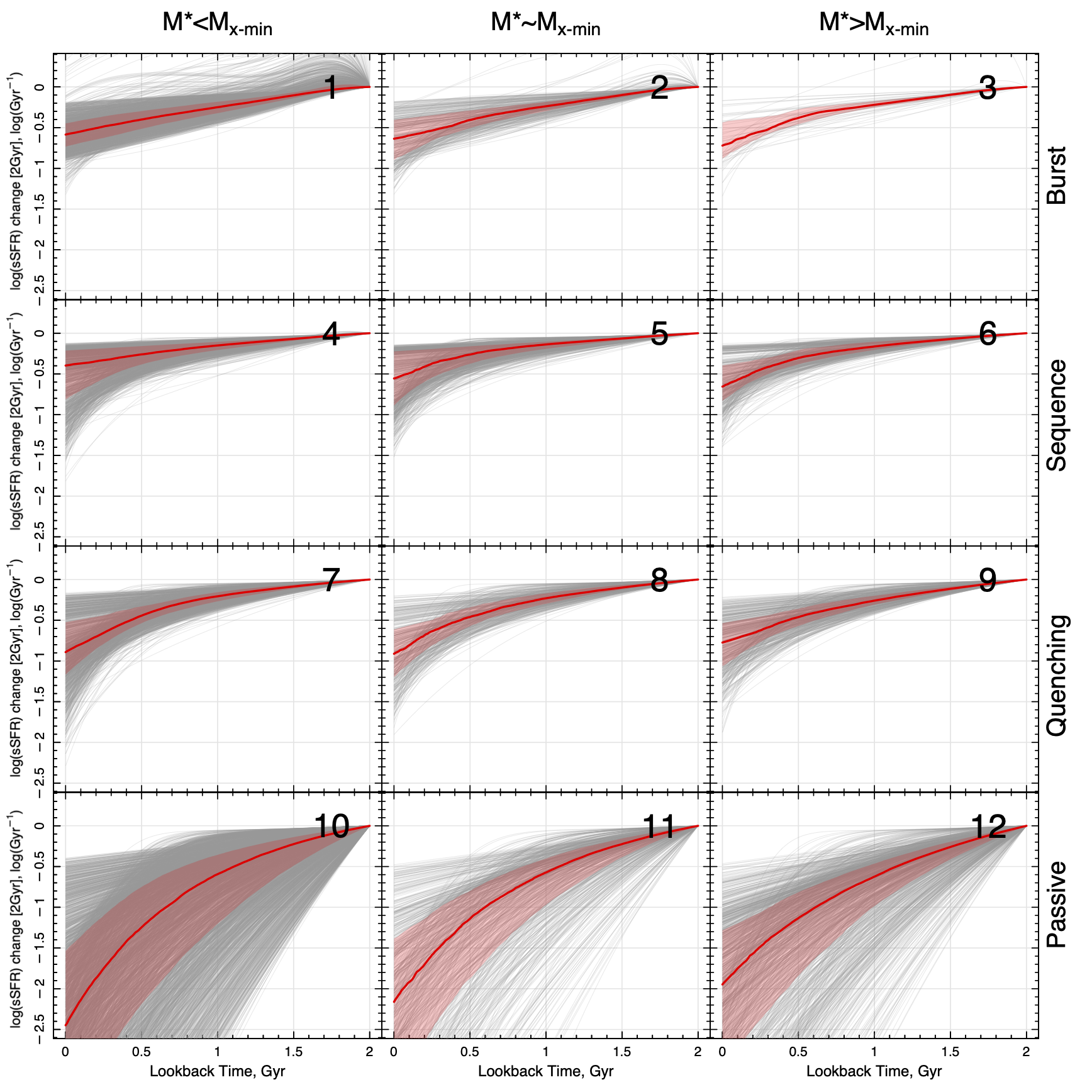}
\caption{The same as Figure \ref{fig:z3_rel}, but now showing SFHs as log$_{10}$(sSFR) note that units are now in Gyr$^{-1}$ for ease of plotting. We still see strong variation of SFHs across the different regions, selected with respect to position within the  sSFR-M$_{\star}$ plane.  }
\label{fig:log_specific_SFHs_z3Rel}
\end{center}
\end{figure*}

\textcolor{black}{Finally, we also repeat the process using log$_{10}$(sSFR) in Figure \ref{fig:log_specific_SFHs_z3Rel}. This displays a combination of the trends seen in the previous two figures. However, the key point we once again indicate, is that there appears to be a stellar mass trend for galaxies along the SFS, with higher stellar mass galaxies typically showing stronger declines in their sSFH than the lower stellar mass counter-parts. We particularly highlight this in each panel as it motivates the choice to split the SFS into two stellar mass ranges when aiming to identify new regions with common SFHs. $i.e.$ this is not just a consequence of using linear SFHs, but there is also some qualitative motivation for this when using these other representations of the recent SFH.  }

\vspace{2mm}

\begin{figure*}
\begin{center}
\includegraphics[scale=0.45]{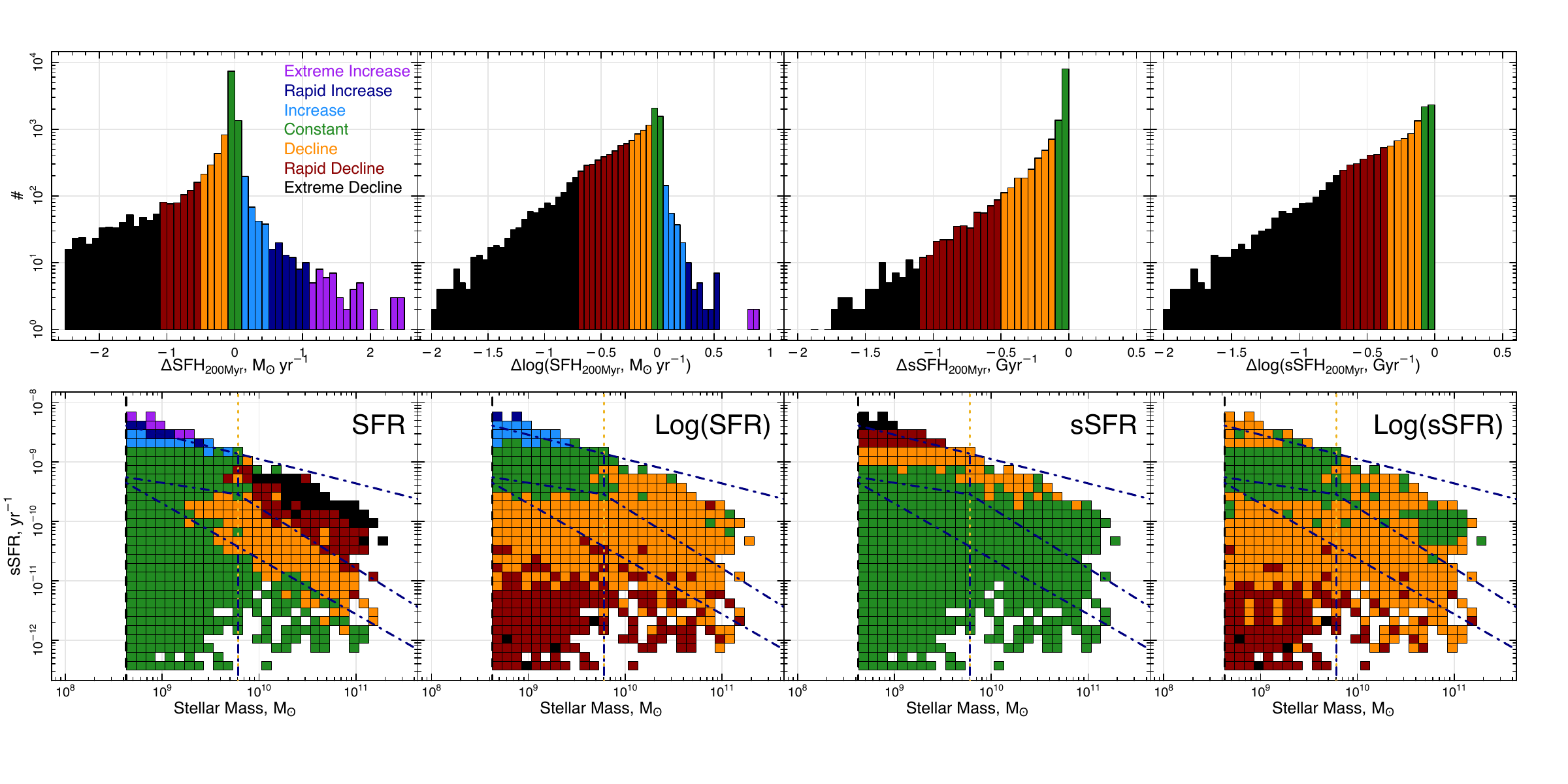}
\caption{The selection of ranges in $\Delta$\,SFH values for each of the different types of SFH for our sample at $0.4<z<0.55$ (top row) and how these $\Delta$\,SFH values vary across the sSFR-M$_{\star}$ plane - similar to Figure \ref{fig:commonBins}.}
\label{fig:slopeRanges}
\end{center}
\end{figure*}

\begin{figure*}
\begin{center}
\includegraphics[scale=0.45]{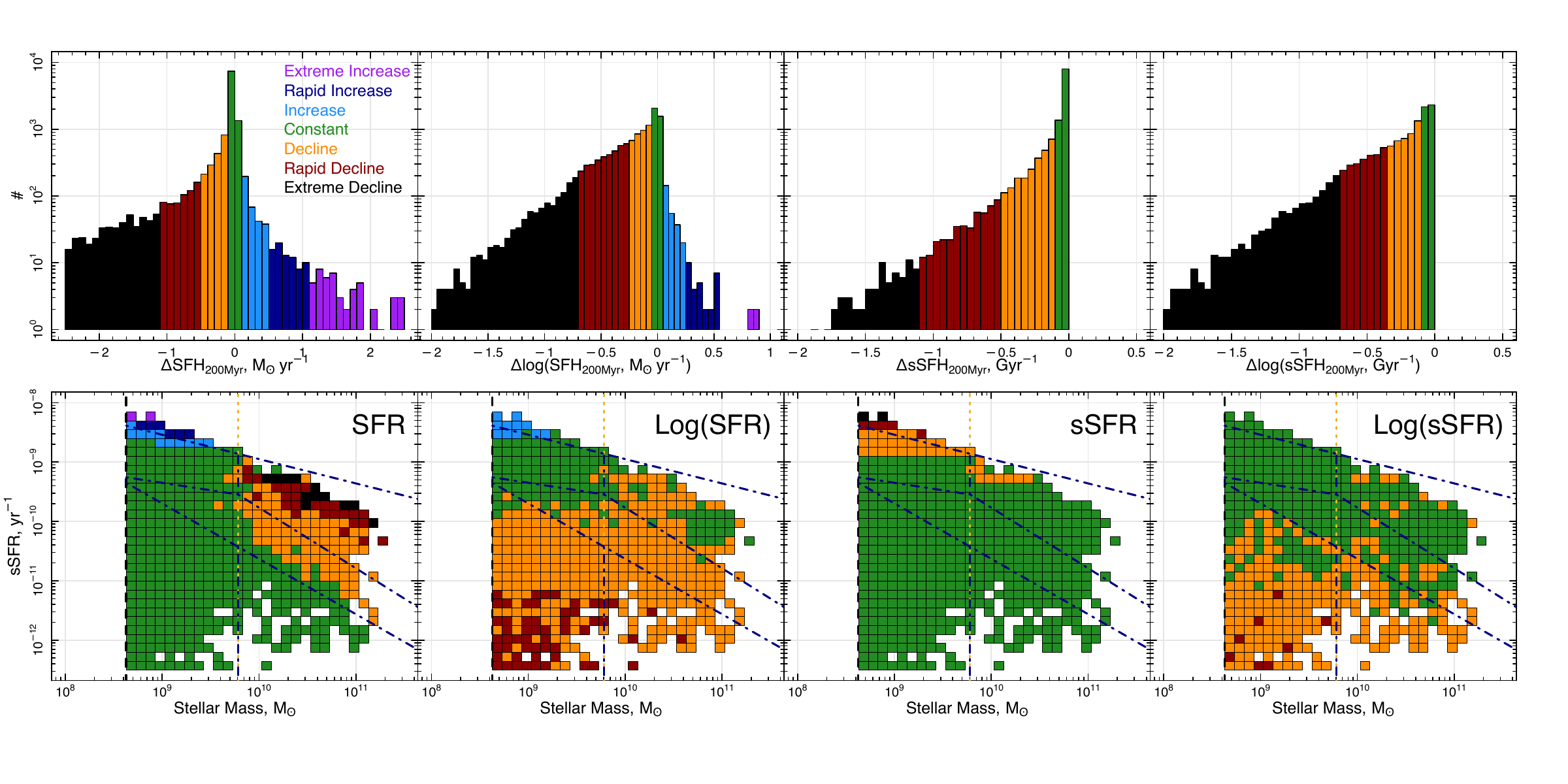}
\caption{The same as Figure \ref{fig:slopeRanges}, but using a 100\,Myr $\Delta$\,SFH time step.}
\label{fig:slopeRanges100}
\end{center}
\end{figure*}

\textcolor{black}{Following this, we next consider our metric for recent change in SFH ($\Delta$SFH) using these other SFH scalings. To do this we define change in star formation over the last 200\,Myr in log$_{10}$(SFR), linear sSFR and in log$_{10}$(sSFR) using the same parametrisation as equation \ref{eq:slope}. We then use this parametrisation to select ranges of different $\Delta$\,SFH values (as is done in the main body of the paper) but for each SFH scaling independently. Due to the differing $\Delta$SFH values, we must use different ranges for each SFH scaling. These are shown for a single epoch at $0.4<z<0.55$  in the top of Figure \ref{fig:slopeRanges}. using these colour-codings we then repeat the process outlined for Figure \ref{fig:commonBins}, and show the most typical $\Delta$\,SFH range in bins across the sSFR-M$_{\star}$ plane (bottom row). Overlayed are the selections used to define new regions with common SFHs (described in the main body of this paper). The trends here are largely consistent with the result displayed above for the SFHs. We also reproduce the same figure but using $\Delta$SFH over the last 100\,Myr to define the recent change in SFH, and show that the overall trends do not change (Figure \ref{fig:slopeRanges100}).}

\vspace{2mm}

\textcolor{black}{The key points we would like to strongly highlight here are that, either using linear SFR, log$_{10}$(SFR) or log$_{10}$(sSFR), some of the overall trends used the define the new selection regions in the main body of the paper, hold true. These are that:} \\

 \textcolor{black}{\noindent $\bullet$ Sources which display $increasing$ recent SFHs (blue/purple points) all sit at low stellar masses and above our top selection line when using both linear and log$_{10}$(SFHs). As such, the region used to define these sources would largely not change when using either SFH scaling.} \\
 
 \textcolor{black}{\noindent $\bullet$ The SFS region shows two distinct populations in terms of typical $\Delta$\,SFH values at low and high stellar masses when using both linear SFHs and log$_{10}$ (SFH) (and potentially log$_{10}$(sSFR)), and that the dividing point between these populations occurs at $\sim$ the minimum dispersion point (vertical gold dashed line). This means that the new regions defined in the main body of the paper, still separate these two populations irrespective of choice of SFH scaling. While the linear and log$_{10}$ SFHs differ as to how extreme the decline in SFH is for the high stellar mass end of the SFS, they still show two different populations - motivating the choice to separate them as such in the main body of the paper.} \\
 
 \textcolor{black}{\noindent $\bullet$ Below the SFS region, galaxies show moderately declining SFHs (orange bins) in both linear SFH and log$_{10}$(SFH), and the current selection boundaries separate this population from the SFS above.}\\ 
 
\textcolor{black}{ \noindent $\bullet$ Below this is where the typical $\Delta$\,SFH values between linear SFH and log$_{10}$(SFHs) differ significantly. Linear SFHs show that galaxies have a constant SFHs, while log$_{10}$(SFHs) find galaxies have rapidly declining SFHs. This is not surprising, given that at these low sSFRs, a very small change in linear SFR can be a large change in log$_{10}$(SFHs). $i.e.$ 0.01 to 0.001 is a small change in linear space, but a large change in log space. When discussing populations here in the interpretation of our results, we will discuss both SFH types. }\\

\textcolor{black}{The key difference between the linear and log SFHs, is that when using log$_{10}$(SFR), there is a very large diversity in SFH slopes at the low stellar mass end and at a fixed stellar mass, which is not seen in the linear SFHs. This diversity is strongly correlated with sSFR, suggesting that at a fixed stellar mass it is correlated with SFR. It is still not obviously clear which metric for recent change in SFH is best to describe this population, and as such, we will highlight both trends when discussing the implication of our results. However, given this is the population that leads to a larger scatter in the SFS at these stellar masses, it is important to caveat any discussion with these differing interpretations.  }

\begin{figure*}
\begin{center}
\includegraphics[scale=0.2]{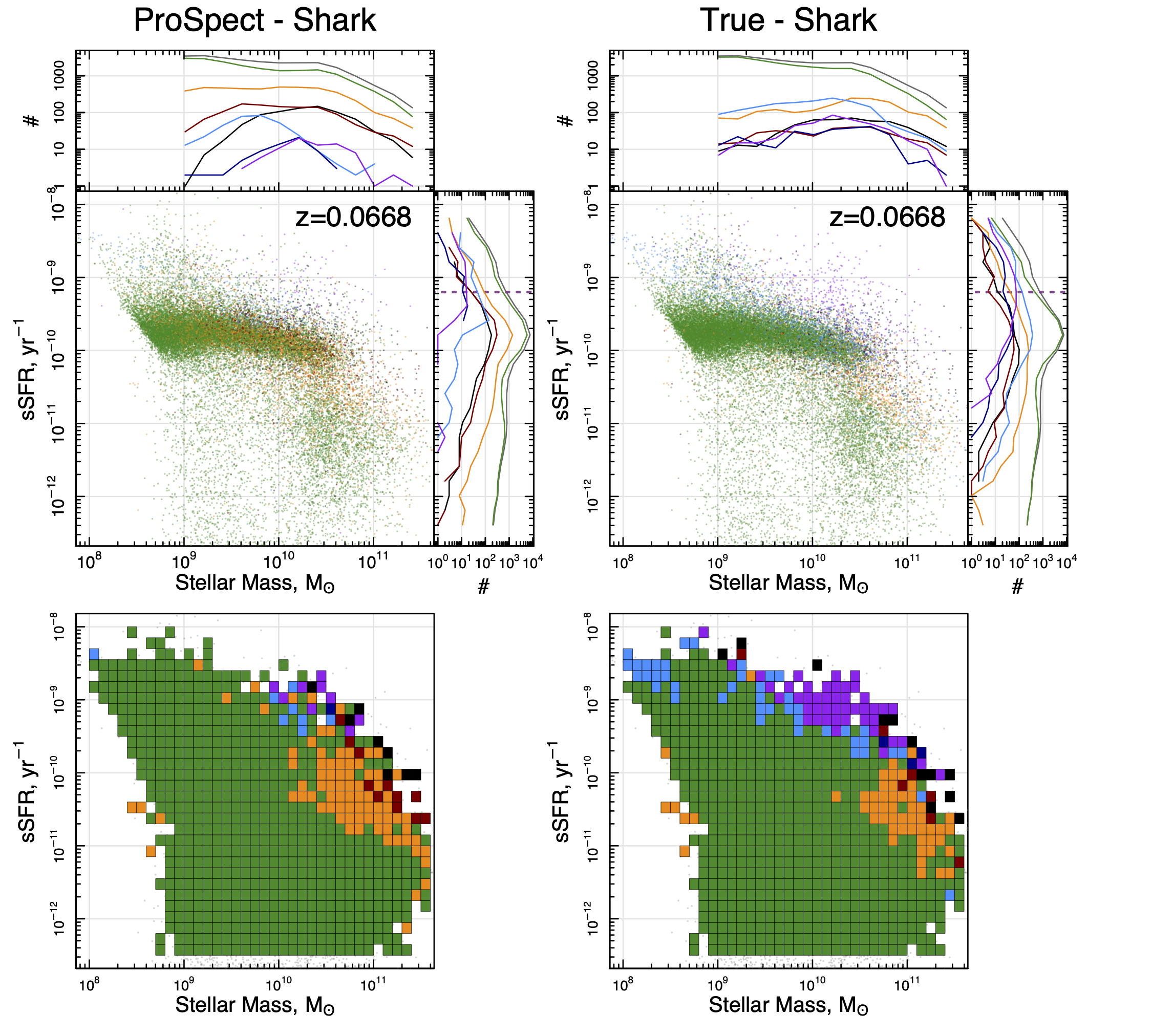}

\caption{Comparison of \textsc{shark} $\Delta$SFH$_{200\mathrm{\,Myr}}$ values across the sSFR-M$_{\star}$ plane derived from both a \textsc{ProSpect} analysis of \textsc{shark} galaxies (left), and the true \textsc{shark} SFH (right). Figures are produced and colour-coded in an identical manner to the DEVILS data in Figures \ref{fig:slope} and \ref{fig:commonBins}. This figure aims to display how well \textsc{ProSpect} can recover the true $\Delta$SFH$_{200\mathrm{\,Myr}}$ values across the sSFR-M$_{\star}$ plane.  Overall, the distributions are largely the same, particularly at the high stellar mass end. However, \textsc{ProSpect} fails to recovers all of the 'increasing' SFH population and biases some constant SFH sources to appear like they have declining SFHs (see text for details). }
\label{fig:shark}
\end{center}
\end{figure*}

\begin{figure*}
\begin{center}
\includegraphics[scale=0.58]{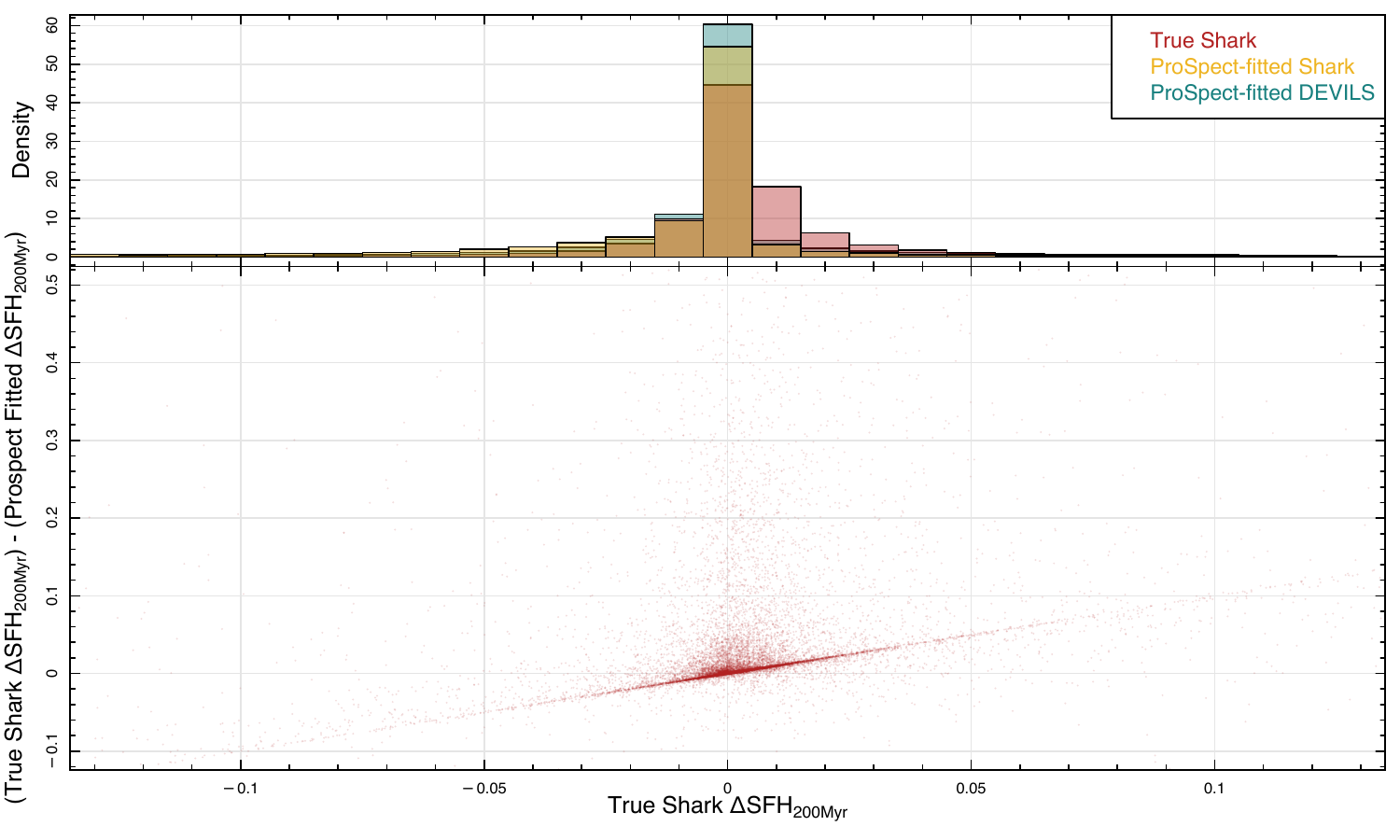}
\includegraphics[scale=0.6]{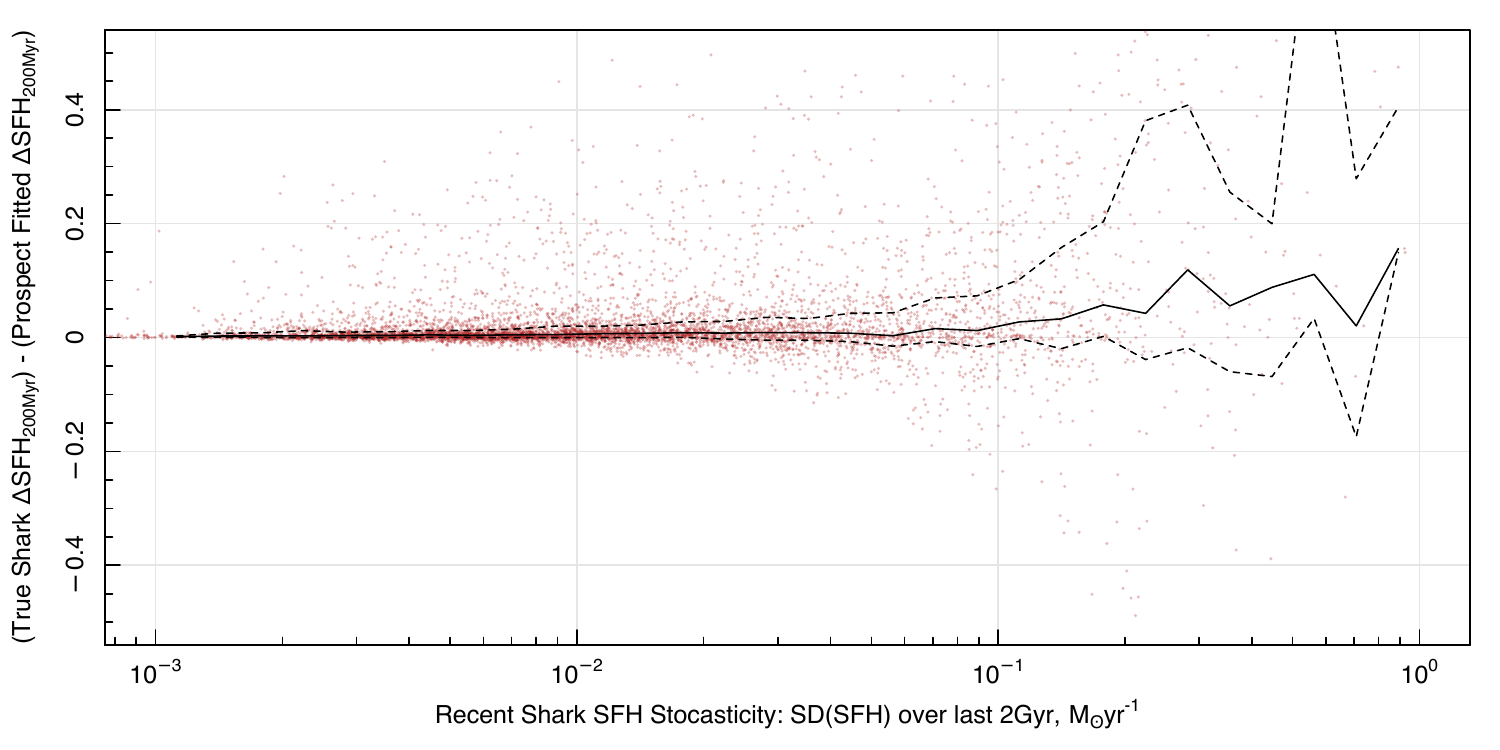}

\caption{ \textcolor{black}{Difference between true \textsc{shark} $\Delta$SFH$_{200\mathrm{\,Myr}}$ values and \textsc{ProSpect}-derived $\Delta$SFH$_{200\mathrm{\,Myr}}$ values as a function of true \textsc{shark} $\Delta$SFH$_{200\mathrm{\,Myr}}$ (top) and the recent SFH stochasticity (bottom) - as traced by the standard deviation in the \textsc{shark} SFH over the last 2\,Gyr. This figure shows the difference between true and fitted $\Delta$SFH$_{200\mathrm{\,Myr}}$  for all M$_{\star}$ $<10^{9.5}$\,M${\odot}$ galaxies in the \citet{Bravo22, Bravo23} analysis. Above the top panel is a histogram of $\Delta$SFH$_{200\mathrm{\,Myr}}$ values in the true \textsc{shark} sample (red), the \textsc{ProSpect}-fitted \textsc{shark} sample (gold) and the observed DEVILS sample used in this work (green). The solid and dashed lines in the bottom panel displays the running median and interquartile range respectively. In the top panel, we see that (for low stellar mass galaxies) \textsc{ProSpect} does appear to miss some galaxies which are increasing in star-formation in \textsc{shark} - this may also be present in the DEVILS observations). In the bottom panel we find that this population is most likely found at the high stochasticity end of the  \textsc{shark} sample - which we had previously noted \textsc{ProSpect} may fail to accurately model.  We find that there is no systematic bias in the \textsc{ProSpect} measurement of $\Delta$SFH$_{200\mathrm{\,Myr}}$ over a range of different \textsc{shark}-simulated galaxies with varying degrees of stochasticity, suggesting that \textsc{ProSpect} can recover recent changes in SFH for the bulk of the low stellar mass populations. For completeness, we also repeat this figure for high stellar mass galaxies and found \textsc{ProSpect} recovers the \textsc{shark} SFH well across all galaxy types. }}
\label{fig:stochasticity}
\end{center}
\end{figure*}

\section{$\Delta$SFH$_{200\mathrm{\,Myr}}$ values in the \textsc{Shark} semi-analytic model}
\label{sec:shark}

To explore the validity of using the $\Delta$SFH$_{200\mathrm{\,Myr}}$ values derived from \textsc{ProSpect} to select galaxies with common recent SFHs, we utilise the \textsc{shark} semi-analytic model \citep{Lagos18, Lagos24}. In a recent study by \cite{Bravo22, Bravo23}, they fit the simulated photometry from 0.05<z<0.1 \textsc{shark} galaxies with \textsc{ProSpect} to explore the colour evolution of galaxies. Here we take the \textsc{ProSpect} fits to the \textsc{shark} galaxies from \cite{Bravo22, Bravo23} and calculate $\Delta$SFH$_{200\mathrm{\,Myr}}$ in an identical manner to our observational data. We then take the true SFH from the \textsc{shark} modelled galaxies and determine the change in star-formation over the last simulation time-step ($\sim$225\,Myr). We note, that we can not replicate this approach at higher redshifts (matched to the DEVILS sample) as the \cite{Bravo22, Bravo23} work only fits local \textsc{shark} galaxies with \textsc{ProSpect}. We also highlight, that there are a number of caveats to be taken into account when directly comparing the \textsc{shark} \textsc{ProSpect} fits to \textsc{ProSpect} measurements of observational data. These are described at length in \cite{Bravo22, Bravo23}, and as such we do not discuss them here. However, we note that the analysis in this section is purely aimed at comparing the $\Delta$SFH$_{200\mathrm{\,Myr}}$ from \textsc{ProSpect} and the true $\Delta$SFH$_{200\mathrm{\,Myr}}$ both derived from \textsc{shark} to asses the validity of the \textsc{ProSpect} values and their application to the observational data.             

In the top row of Figure \ref{fig:shark} we show the (true) sSFR-M$_{\star}$ plane for \textsc{shark} galaxies in a similar manner to Figure \ref{fig:slope}. In the left column points are colour-coded by the \textsc{ProSpect}-fitted $\Delta$SFH$_{200\mathrm{\,Myr}}$ value, and in the right column points are colour-coded by the true change in \textsc{shark} SFH. In the bottom row we show the binned median SFH version of these figures as in Figure \ref{fig:commonBins}. Colour ranges in all panels are identical to those previously used in the DEVILS sample.

Encouragingly, the first thing we see from these figures is that quantitatively the overall distributions of $\Delta$SFH$_{200\mathrm{\,Myr}}$ values across the sSFR-M$_{\star}$ plane are very similar between the two methods, and in fact very similar to the observational trends we see in DEVILS. This is particularly evident in the bottom panels, there the regions dominated by particular $\Delta$SFH$_{200\mathrm{\,Myr}}$ values are common across both methodologies. This suggests that for the average trends across the sSFR-M$_{\star}$ plane, the \textsc{ProSpect} $\Delta$SFH$_{200\mathrm{\,Myr}}$ values do capture the true recent SFH. This is also found to be consistent with the conclusions of \cite{Bravo23} when comparing the \textsc{shark} \textsc{ProSpect} predictions and the true \textsc{shark} evolutionary trends. As such, to first order, using the $\Delta$SFH$_{200\mathrm{\,Myr}}$ derived from \textsc{ProSpect} to explore the observational trends in DEVILS and select objects with common recent SFHs is likely valid.      

However, as noted previously care must be taken when using \textsc{ProSpect} to determine the recent SFH of individual sources. As expected, the \textsc{ProSpect} analysis does fail to determine the true $\Delta$SFH$_{200\mathrm{\,Myr}}$ value for many sources with increasing SFHs (blue and purple points). The \textsc{ProSpect} analysis does find these types of sources, and they do lie in similar regions of the sSFR-M$_{\star}$ plane. However, they are found in far fewer numbers than the true SFHs would suggest. 

\textcolor{black}{Taking this a stage further, we can also consider how well \textsc{ProSpect} can recover the $\Delta$SFH$_{200\mathrm{\,Myr}}$ values for bursty low mass galaxies, which likely have rapidly changing SFHs. Figure \ref{fig:stochasticity} hows the difference between true \textsc{shark}-simulated  and \textsc{ProSpect}-fitted $\Delta$SFH$_{200\mathrm{\,Myr}}$ values for all M$_{\star}$ $<10^{9.5}$\,M${\odot}$ galaxies in the \cite{Bravo22, Bravo23} analysis, as a function of}  \textcolor{black}{both true \textsc{shark}-simulated $\Delta$SFH$_{200\mathrm{\,Myr}}$ and recent stochasticity in the true \textsc{shark} SFH. This stochasticity is traced by the standard deviation in the SFH over the last 2\,Gyr. This figure shows that firstly, \textsc{ProSpect} may fail to recover the correct $\Delta$SFH$_{200\mathrm{\,Myr}}$ value for some galaxies with increasing SFHs (as evidenced by the paucity of object in both the \textsc{ProSpect}-fitted \textsc{shark} data and DEVILS data in comparison to true shark values at positive $\Delta$SFH$_{200\mathrm{\,Myr}}$). In the bottom panel we find that this is likely due to highly stochastic sources (which we have previously noted that \textsc{ProSpect} may fail to capture). Secondly, we find that there is     no strong biases in the \textsc{ProSpect}-fitted $\Delta$SFH$_{200\mathrm{\,Myr}}$ values across a broad range of stochasticities in\textsc{shark}. This suggests that \textsc{ProSpect} can recover recent changes in SFH for the majority low stellar mass populations. However, care must be taken as there are clearly some limitations of \textsc{ProSpect} in this low stellar mass regime. For completeness, we note that we replicate Figure \ref{fig:stochasticity} for M$_{\star}$ $>10^{10}$\,M${\odot}$ galaxies and find that this bias is removed - suggesting this largely only impacts low stellar mass galaxies. }

Interestingly, the \textsc{ProSpect} analysis also finds more galaxies with slowly declining SFHs (orange points) at low stellar masses (M$_{\star}$ $<10^{10.0}$\,M${\odot}$) and on the SFS. The slowly declining populations at higher stellar masses are comparable and take up a similar region of the sSFR-M$_{\star}$ plane. While at lower stellar masses the \textsc{ProSpect} analysis finds a significant fraction of the SFS population has a declining SFH, which is not evident in the true SFH. This is likely a artefact of  \textsc{ProSpect} fitting a skew log-normal SFH to all systems. In the case of a flat SFH over all epochs, the skew log-normal distribution must fit a peak in star-formation intensity at t$_{\mathrm{lookback}}>0$, and then a decline in star-formation following this. This would then bias a sample of relatively flat SFH sources to have slightly declining SFHs.  However interestingly, this population does not appear to be evident in the DEVILS observational data, where we do not see a significant fraction of systems with slowly declining SFHs on the SFS. This indicates that care must be taken when selecting galaxies based on small differences in $\Delta$SFH$_{200\mathrm{\,Myr}}$ values, as the \textsc{ProSpect} analysis may drive small biases towards moderately declining SFHs. This once again, indicates that the \textsc{ProSpect} SFHs should only be used to explore general trends across the sSFR-M$_{\star}$ plane, not individual sources- as we do in this work.

In combination, this \textsc{shark} analysis suggests that \textsc{ProSpect} SFHs (and the derived $\Delta$SFH$_{200\mathrm{\,Myr}}$ values), are adequate to explore the overall evolution of galaxies in subregions of the sSFR-M$_{\star}$ plane, but are likely not sensitive to short time-scale fluctuations in SFH for individual sources; as previously discussed.

\section{Evolution of SFR-M$_{\star}$}

For completeness, here we also show a similar Figure to Figure \ref{fig:PlaneEvol}, but for SFR instead of sSFR. We reproduce this figure here such that readers who prefer to work in the SFR-M$_{\star}$ plane can better understand our results in terms of the evolution of the SFS.

\begin{figure*}
\begin{center}
\includegraphics[scale=0.13]{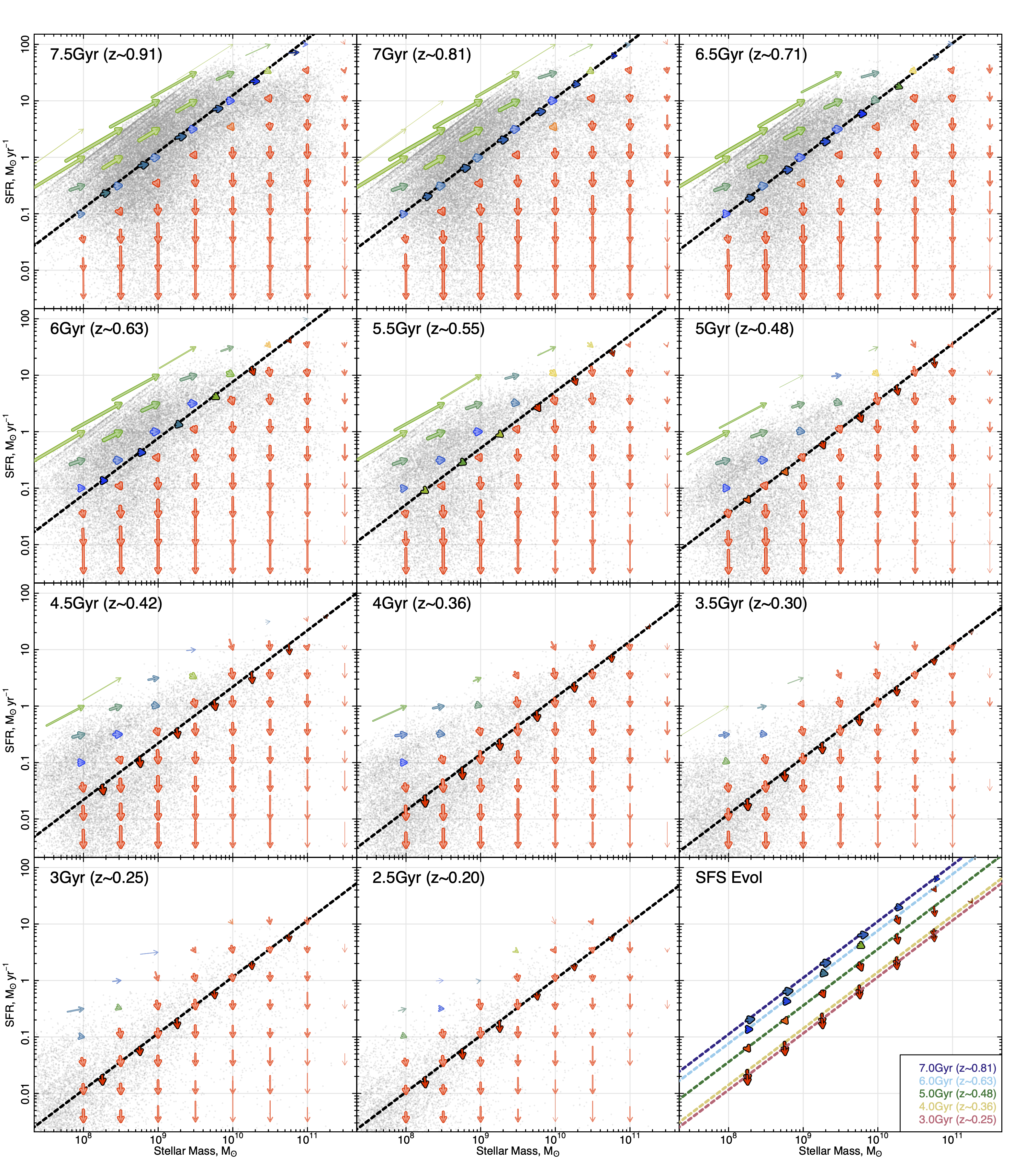}

\caption{The same as \ref{fig:PlaneEvol}, but showing the SFR-M$_{\star}$ plane. This figure is reproduced in this manner for ease of understanding for the reader. In this plane, galaxies with a constant SFR move horizontally. Between $\sim$6\,Gyr and $\sim$4\,Gyr lookback time the SFS transitions from moving horizontally to the right, to vertically down.  }
\label{fig:PlaneEvolSFR}
\end{center}
\end{figure*}

\bsp	
\label{lastpage}
\end{document}